\documentclass[12pt]{article}

\usepackage{amsmath}
\usepackage{amssymb}
\usepackage{slashed}
\usepackage[numbers,sort&compress]{natbib}
\usepackage{hyperref}
\usepackage{cleveref}
\crefname{equation}{Eq.}{Eqs.}
\crefname{figure}{Fig.}{Figs.}
\crefname{table}{Table}{Tables}
\crefname{section}{Section}{Sections}

\usepackage[letterpaper,margin=1in,bottom=1in]{geometry}
\usepackage{float} 
\usepackage{parskip} 
\usepackage{tabulary} 
\usepackage{color} 
\usepackage{soul} 
\usepackage{subfigure}
\usepackage{graphicx}
\usepackage[section]{placeins} 

\def\lhc2{LHC~Run~II}

\usepackage{cleveref}

\newcommand{\code}[1]{\texttt{#1}}
\newcommand{\GeV}{~\textrm{GeV}}

\def\co{coannihilation~}

\def\.4{\vspace{-.5cm}}
\newcommand{\ifb}{~\textrm{fb}^{-1}}

\def\beq{\begin{equation}}
\def\be{\begin{equation}}
\def\beqn{\begin{eqnarray}}
\def\ee{\end{equation}}
\def\eeq{\end{equation}}
\def\eeqn{\end{eqnarray}}

\def\co{coannihilation~}

\def\co{coannihilation~}
\def\G{\tilde G}
\def\mP{M_{\rm Pl}}

\author{
Amin Aboubrahim\footnote{Email: a.abouibrahim@northeastern.edu}~\ and 
Pran Nath\footnote{Email: p.nath@northeastern.edu}\\~\\
Department of Physics, Northeastern University,
Boston, MA 02115-5000, USA
}

\title{Supergravity Models with 50-100 TeV Scalars, SUSY Discovery at  the LHC and Gravitino Decay Constraints}

\begin{document}
\maketitle
\date

\textbf{Abstract: } 
We investigate the possibility of testing supergravity unified models with scalar masses in the range 50-100 TeV
and much lighter gaugino masses at the Large Hadron Collider.  The analysis is carried out under the constraints that 
models  produce the Higgs boson mass consistent with experiment and also produce dark matter consistent with WMAP and PLANCK
 experiments.  A set of benchmarks  in the supergravity parameter space are  investigated using a combination of signal regions which are optimized  for the model set. It is found that some of the models
 with scalar masses in the 50-100 TeV mass range are discoverable with as little as 100 fb$^{-1}$ of integrated luminosity and
 should be accessible at the LHC RUN II. The remaining  benchmark models are found to be discoverable with less
 than 1000 fb$^{-1}$ of integrated luminosity and thus testable in the high luminosity era of the LHC, i.e., at HL-LHC. 
  It is  shown that  scalar masses in the 50-100 TeV range  but gaugino masses much lower in mass 
 produce  unification of gauge coupling constants, consistent with experimental data at low scale, 
 with as good an accuracy (and sometimes even better) as 
 models with low ($\mathcal{O}(1)$ TeV) weak scale supersymmetry. 
Decay of the gravitinos for the supergravity model benchmarks are investigated and it is shown that they decay before the Big Bang Nucleosynthesis (BBN). 
Further, we investigate the non-thermal production of neutralinos from gravitino decay and it is found that the non-thermal 
contribution to the dark matter relic density is negligible relative to that from the thermal production of neutralinos 
for reheat temperature after inflation up to $10^9$ GeV. 
  An analysis of the direct detection of dark matter for SUGRA models with high scalar masses is also discussed. 
  SUGRA models with scalar masses in the range 50-100 TeV have several  other attractive features such as they help alleviate the SUSY CP problem  and help suppress proton decay from baryon and lepton number violating  dimension five operators.

\newpage

\section{Introduction}\label{sec:intro}

Supersymmetry (SUSY) has not been observed thus far, which implies that the weak supersymmetry scale is higher than 
was expected before the Higgs boson~\cite{Englert:1964et, Higgs:1964pj, Guralnik:1964eu} 
was discovered at the Large Hadron Collider (LHC) 
 and specifically before the measurement of its mass at $\sim 125$ GeV~\cite{Chatrchyan:2012ufa, Aad:2012tfa}.
Analyses within high-scale supergravity grand unified models (SUGRA)~\cite{msugra} (for a review see~\cite{book}) 
show that the loop correction to the Higgs boson mass in supersymmetry must  be sizable, which in turn implies 
a value of weak SUSY scale lying in the several TeV region~\cite{Akula:2011aa,Arbey:2012dq,susy-higgs,Baer:2015fsa}. There is another constraint that explains the possible reason for the lack of detection of a supersymmetric
 signal. In supergravity grand unified models with R-parity conservation, neutralino is the lightest supersymmetric particle  (LSP) over most of the parameter space of models~\cite{Arnowitt:1992aq} 
 and thus a candidate for dark matter. The annihilation of the neutralino in sufficient amounts to have its relic density consistent with the WMAP~\cite{Larson:2010gs} and the PLANCK~\cite{Ade:2015xua} experimental results imposes additional constraints. 
  Specifically if the neutralino is bino-like, one needs coannihilation (for early work see~\cite{Griest:1990kh})
 to have consistency with experiment. 
 However, \co implies that the next-to-lightest supersymmetric particle (the NLSP) must be close
 to the LSP with a small mass gap to ensure efficient annihilation of the LSP. The existence of the small mass gap in turn implies
 that the final states in the decay of the NLSP will be soft making them difficult to detect.  
 
 Coannihilation appears in supergravity models
 with universal as well as with non-universal boundary conditions at the grand unification scale which lead to 
  a large sparticle landscape~\cite{Feldman:2007zn}.   The large landscape  
 includes non-universalities in the gaugino sector~\cite{Ellis:1985jn,nonuni2} and in the matter and 
Higgs sectors~\cite{NU}. 
As mentioned above the measurement of the Higgs boson mass at $125$ GeV, implies that the scale of weak scale
supersymmetry lies in the several TeV region.
 Assuming universality of 
 the scalar mass at the GUT scale a high value of the universal scalar mass $m_0$ is indicated.  
Quite interestingly  it has been previously argued that scalar masses  could  be large and natural on the hyperbolic branch  
of radiative breaking of the electroweak symmetry~\cite{Chan:1997bi,Feng:1999mn,Chattopadhyay:2003xi,Baer:2003wx,Feldman:2011ud,Akula:2011jx,Ross:2017kjc}.  In this analysis we consider much higher values of scalar masses than typically
considered in supergravity models, i.e., scalar masses lying in the 50-100 TeV region.  This regime of scalar masses
  help alleviate some of the problems associated with low values of the weak SUSY scale 
   such as the SUSY CP problem (see, e.g.,~\cite{Ibrahim:2007fb})  
    and fast proton decay from baryon and lepton number violating 
        dimension five operators~\cite{Nath:2006ut,Liu:2013ula}.
  Further, in supergravity unified models the gravitino mass $m_{\tilde G}$ and the scalar mass $m_0$ are related  and thus 
   a large $m_0$ in the  50-100 TeV range helps alleviate the  problem arising from the late decay of the gravitino which would upset the Big Bang Nucleosynthesis (BBN).

   The search for supersymmetric signatures in models of the above type with high values of $m_0$ would necessarily focus on light gauginos and a compressed spectrum. Models with \co and a compressed spectrum have been analyzed 
   over the years in a variety of settings involving  chargino, stau, stop and gluino \co 
  (For some recent works  on \co  related to the analysis here see 
\cite{Kaufman:2015nda, Nath:2016kfp, Aboubrahim:2017aen}. For other related works see
\cite{Ellis:1998kh,Nihei:2002sc,Baer:2005jq,Arnowitt:2008bz,Feldman:2009zc,Akula:2013ioa,Camargo-Molina:2013sta,Kowalska:2015zja,Florez:2016lwi}.
For recent theory papers related to supersymmetry and compressed spectrum see~\cite{Dutta:2015exw,Berggren:2015qua,Berggren:2016qjh,LeCompte:2011fh}
and for experimental searches for supersymmetry with a compressed spectrum see~\cite{Khachatryan:2016mbu,Khachatryan:2016pxa,MORVAJ:2014opa}).

 The outline of the rest of the paper is as follows: In section 2 we discuss a set of benchmarks for supergravity models  with scalar masses in the 50-100 TeV mass range. The benchmarks are chosen so they satisfy the radiative breaking of electroweak symmetry,  give a Higgs boson mass consistent with experiment, and produce a relic density for neutralino cold dark matter consistent with WMAP and PLANCK. 
{In section 3,  we compare the gauge coupling unification for supergravity models with low weak scale supersymmetry
vs high weak scale supersymmetry consistent with the LEP data.
 In section 4, we discuss the production of supersymmetric particles and their decays. 
  Here we exhibit the  cross sections for the production of the SUSY particle pairs $\tilde\chi_2^0\tilde\chi_1^\pm$, $\tilde\chi_1^+\tilde\chi_1^-$, $\tilde\chi_1^0\tilde\chi_1^{\pm}$ and $\tilde{g}\tilde{g}$.  The sparticles decay with a neutralino and standard model (SM) particles in the final states. The signature analysis of these requires a knowledge of the backgrounds arising from the production and decay of the standard model particles.  Here we use the backgrounds
     published by the SNOWMASS group~\cite{Avetisyan:2013onh}.    
      Section 5 is devoted to the signature analysis of the benchmarks and an
  analysis of the minimum integrated luminosity needed at the LHC operating at $14$ TeV
 for the $5\sigma$ discovery.  Here a comparison of the different signature regions is also 
 made and combined signal region results are exhibited where models are arranged in terms of
 ascending order in the minimum integrated luminosity needed for a $5\sigma$ discovery.
  In section 6, we discuss the gravitino decay and its possible contribution to the LSP relic density. 
 In section 7, we discuss direct detection of dark matter for SUGRA models with 50-100 TeV scalars.  
 Conclusions are given in section 8.
 
\section{SUGRA models with 50-100 TeV scalar masses }
  To analyze supergravity models with scalar masses in the range 50-100 TeV,  we need to explore the supergravity 
  parameter space consistent with radiative breaking of the electroweak symmetry, the Higgs boson mass and relic density constraints.  We are also interested in exploring the parameter space where the gaugino masses are relatively light with masses that would be accessible at the LHC.  Further, we limit ourselves to the case that  R-parity is 
  conserved so that the LSP is stable.
   Often in most of the parameter space of supergravity models it is found that  under constraints of radiative breaking of the electroweak symmetry  the lightest neutralino is the LSP~\cite{Arnowitt:1992aq},  and under the assumption of R-parity conservation, it is a candidate for dark matter. In this case the constraints on dark matter relic density given by WMAP and PLANCK become relevant. It is found that in part of the parameter space, where the Higgs boson mass constraint is satisfied the neutralino turns out to be mostly a bino. The annihilation cross section for the bino-like neutralino is small and thus the neutralinos in the early universe cannot efficiently annihilate themselves to standard model particles to produce  the desired relic density. Here one needs \co to reduce the neutralino relic density to be compatible with WMAP and PLANCK data on cold dark matter. 
   
    Coannihilation can be easily achieved in supergravity models with non-universal gaugino masses. 
   One such possibility is non-universality between the  $U(1)$ gaugino mass $m_1$  and 
    the $SU(2)$ gaugino mass $m_2$. In this case the light 
    chargino $\tilde \chi_1^{\pm}$ may lie close to the LSP neutralino $\tilde \chi_1^0$ which results in \co while the mass of the $SU(3)$ gaugino $m_3$ is relatively much larger, i.e. $m_3 >> m_1, m_2$. 
 The  parameter space of this model is thus given by $m_0\,, A_0\,,m_2<m_1<<m_3\,$, $\tan\beta\,, \text{sign}(\mu)\,$,
where $A_0$ is the universal trilinear scalar coupling at the grand unification scale, $\tan\beta=\langle H_2\rangle/\langle H_1\rangle$, where $H_2$ gives mass to the up quarks and $H_1$ gives mass to the down quarks and the leptons, and sign$(\mu)$ is the sign of the Higgs mixing parameter which enters in the superpotential in the term $\mu H_1 H_2$. Using the above input parameters, the sparticle spectrum is generated using \code{SoftSUSY 4.0.1}~\cite{Allanach:2001kg, Allanach:2016rxd} while the analysis of the relic density is done using \code{micrOMEGAs 4.3.2}~\cite{Belanger:2014vza}. SUSY Les Houches Accord formatted data files are processed using \code{PySLHA}~\cite{Buckley:2013jua}.

To determine the prospects of SUSY discovery for SUGRA models  with high scalar masses,  ten benchmark
points were generated lying in the mass range 50-100 TeV. The benchmarks selected were those satisfying the 
radiative electroweak symmetry breaking constraints (for a review see~\cite{Ibanez:2007pf}), 
Higgs boson mass constraint with  the Higgs boson mass lying in the range $125\pm2$~GeV.
These model points also satisfied the constraint on the relic density of the LSP neutralino so that 
 $\Omega_{\tilde \chi_1^0} h^2<0.128$. These benchmarks  are displayed in Table~\ref{tab1} and the corresponding sparticle masses, the Higgs boson mass and relic density are shown in Table~\ref{tab2}. Note that $A_0 \approx 2m_0$ for all the benchmarks consistent with previous works that the large loop correction to the Higgs boson mass requires a 
substantial $A_0$ (see, e.g.,~\cite{Akula:2011aa}). 
Further, on average, $m_2\approx 0.8 m_1$, which is needed to bring the chargino mass close to the LSP mass.
The mass gap between the LSP mass and the chargino mass lies in the range $\sim 15$ to 28 GeV leading to a compressed spectrum 
for the LSP and the NLSP. The compressed spectrum  implies that the decay of the NLSP will lead to soft leptons and jets making the
detection of supersymmetry a  challenging task for this part of the parameter space.  
 
 The benchmarks of Table 1 are used to generate Table~\ref{tab2}  which exhibits a set of sparticle masses including the light
 spectrum as well as the heaviest squark and the average sfermion mass. 
       The heaviest squark has a mass roughly $m_0$, whereas the average sfermion mass appears to be lower than $m_0$. The reason
       for the average sfermion mass being lower than $m_0$ is due to the 
  presence of  lighter third generation squarks  (see Fig.~\ref{fig1}) where the analysis is done using 
 ~\code{softSUSY}. In the regime of $m_0$ in the range 50-100 TeV, spectrum generators tend to be less stable. This arises from convergence problems mainly due to  $\mu$ becoming highly dependent on the top Yukawa coupling~\cite{Belanger:2005jk}. Using \code{softSUSY}, we have made sure that none of the points considered in this analysis suffer from such a convergence problem. Further, model points have been cross tested using  \code{ISAJET}~\cite{Paige:2003mg} which  gives a spectrum within the same mass range as \code{softSUSY}. We note in passing
 that   the 
parameter set of Table 1 has not been ruled out by experiment, as can be seen by comparing the spectrum of Table \ref{tab2} 
with the $\tilde{\chi}_1^{\pm}$-$\tilde{\chi}^0_1$ or the $\tilde{\chi}_1^{\pm}$-$\tilde{g}$ exclusion plots from experiment~\cite{Aad:2015eda, ATLAS:2017uun}.

 \section{Gauge Coupling Unification With High Mass Scalars}
One of the well known successes of SUSY is that it gives a unification of gauge couplings consistent 
with LEP data~\cite{drw}.
In such analyses the typical assumption made is of sparticle masses in the  sub-TeV-TeV range.  In the analysis 
of Table \ref{tab2} we find a split sparticle spectrum where the gauginos, $\tilde \chi_1^0, \tilde \chi_2^0, \tilde \chi_1^{\pm}, 
\tilde g$ have  low masses while the scalars are 50-100 times larger in mass. It is then of interest to ask if the unification of gauge couplings holds 
to the same degree of accuracy for the models of Table \ref{tab2} as compared to models with all sparticle masses low lying in 
the sub-TeV-TeV range. 
To check this, we plot the running of  $\alpha_i^{-1} (i=1,2,3)$ ($\alpha_i=g_i^2/4\pi$, where $g_i$ are the gauge 
couplings for the gauge groups $U(1)_Y, SU(2)_L, SU(3)_C$ and $g_1=  \sqrt{\frac{5}{3}} g_Y$)
 for model point (j) using \code{softSUSY}.
Fig.~\ref{fig7} shows two plots: The plot  on the left exhibits the running of $\alpha_i^{-1}$ for a small universal scalar mass $m_0=740$ GeV  with all other parameters the same as in Table \ref{tab1} for point (j).  
Defining the GUT scale as the bi-junction where $\alpha_1$
and $\alpha_2$ meet, one finds that $\alpha_3$ misses the bi-junction  by 3.6 \%.  
 The plot  on the right exhibits the running of $\alpha_i^{-1}$ with all parameters the same as in Table \ref{tab1} for point (j).  
Here one finds that $\alpha_s$ misses the bi-junction by 1.7 \%. 
Thus one finds that unification of gauge couplings  occurs to a good accuracy in each case
with the larger $m_0$ case showing a small improvement in this case. A similar result is observed for
other model points of Table \ref{tab1}. Thus we conclude that models with scalar masses in the 50-100 TeV range and gaugino masses much lower produce unification of gauge couplings with about the same degree of  accuracy
as sub-TeV-TeV scale SUSY models and in some cases with a slight improvement. 

\begin{figure}[htp]
\centering
\includegraphics[width=.48\textwidth]{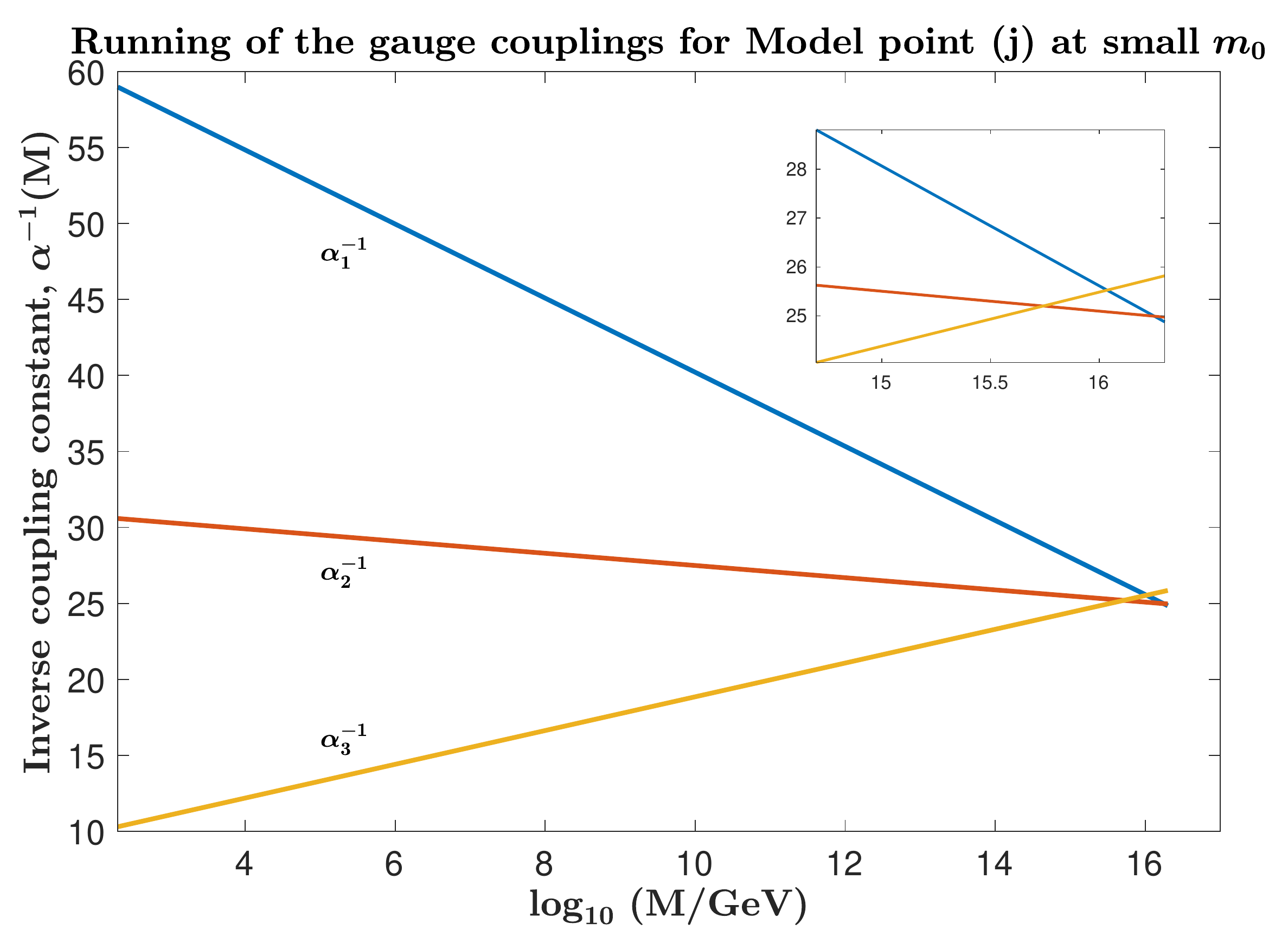}
\includegraphics[width=.48\textwidth]{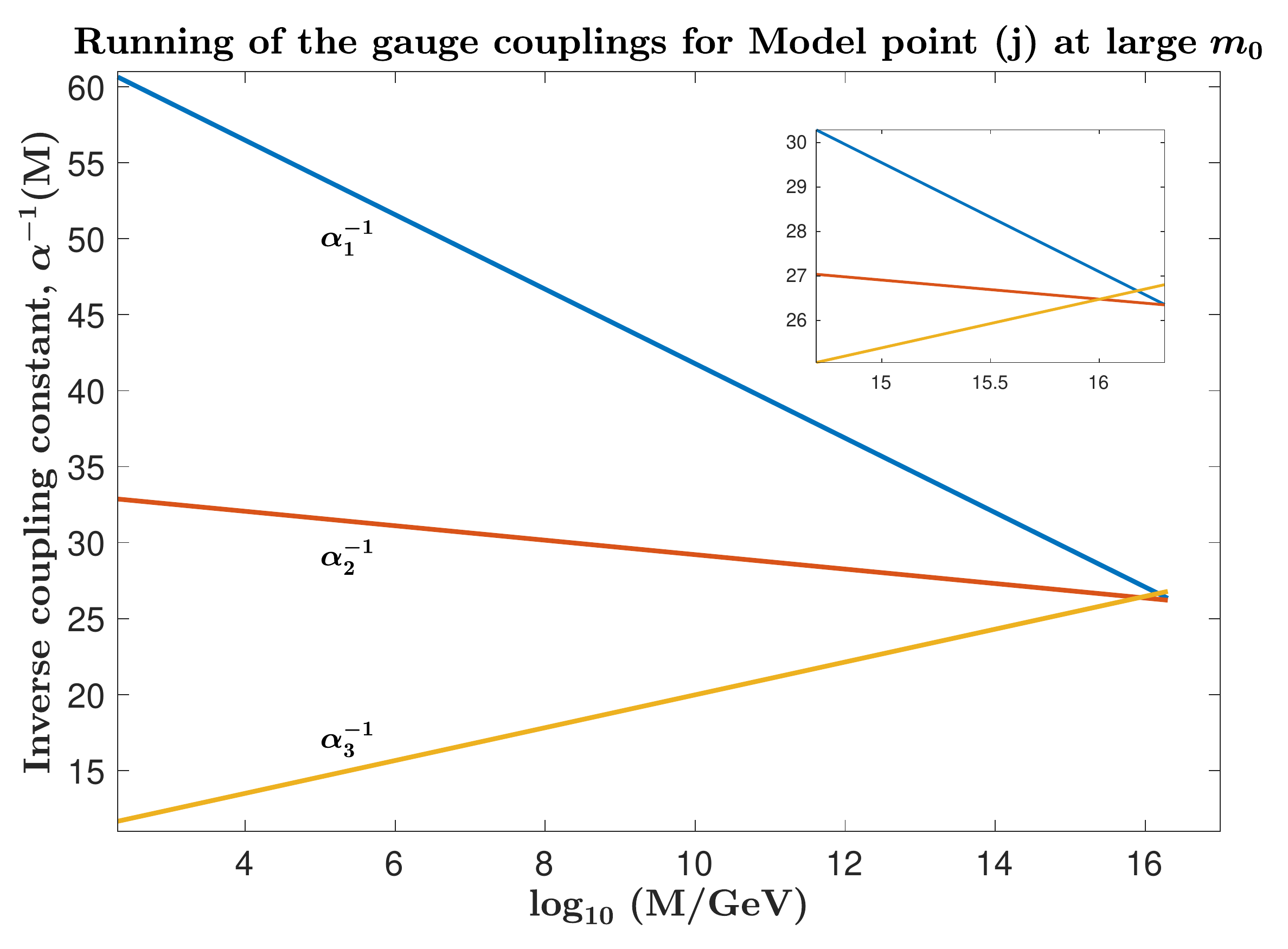}
\caption{A comparison of the unification of gauge couplings  for low scale and high scale SUSY models
for the benchmark (j) of Table \ref{tab1}.
Left panel: the plot shows the running of $\alpha_i^{-1}$  ($i=1,2,3$) for $U(1)_Y, SU(2)_L, SU(3)_C$ 
 for low scale SUSY with universal scalar mass $m_0=740$ GeV and all other parameters the same as for benchmark (j) in Table \ref{tab1},  
 where $\alpha_s$ misses the bi-junction where $\alpha_1$ and $\alpha_2$ meet  by $\sim$ 3.6 \%.
Right panel: the plot shows the running of $\alpha_i^{-1}$ for the high scale SUSY with universal scalar mass $m_0=74$ TeV and all other parameters the same as in the left panel. Here $\alpha_s$ misses the bi-junction by $\sim$ 1.7 \%. Thus in this case the
high scale SUSY shows a small improvement in the unification fit  relative to low scale SUSY.}
\label{fig7}
\end{figure}

\section{LHC sparticle production and decay at 14 TeV}
The benchmarks listed in Table~\ref{tab1} are used in further Monte Carlo analysis in the signal regions appropriate for the detection of a supersymmetric signal at the LHC. 
This analysis was performed with \code{MADGRAPH 2.5.5}~\cite{Alwall:2014hca}. First, the Feynman diagrams were calculated for all possible decays of the form $pp\to\text{SUSY SUSY}$, where ``$\text{SUSY}$" can be any 
$R$ parity odd minimal supersymmetric standard model (MSSM) particle. With the sparticle spectra of the benchmarks calculated by \code{SoftSUSY 4.0.1}, as well as the decay widths and branching ratios computed by \code{SDECAY} and \code{HDECAY} operating within \code{SUSY-HIT}~\cite{Djouadi:2006bz}, \code{MADEVENT} was used to simulate 50,000 MSSM decay events for each benchmark point. Hadronization of resultant particles is handled by \code{PYTHIA8}~\cite{Sjostrand:2014zea} where ISR and FSR jets are switched on, and ATLAS detector simulation and event reconstruction is performed by \code{DELPHES 3.4.1}~\cite{deFavereau:2013fsa}. A large set of search analyses were performed on the generated events for each benchmark point. The analyses used \code{ROOT 6.08.06}~\cite{Antcheva:2011zz} to implement the constraints of the
search region for the signal regions involving leptons, jets and missing transverse energy in the final state. 

To allow comparison to the background, all of the signal region analyses were applied to pre generated backgrounds published by the SNOWMASS group~\cite{Avetisyan:2013onh}. For each benchmark, a calculated implied integrated luminosity allowed direct comparison to the backgrounds. Each individual background process from the SNOWMASS  set was scaled by its own implied integrated luminosity and combined to determine a total background count for each signal region. The various background samples are grouped according to the generated final state, with a collective notation given by
\begin{equation}
\begin{aligned}
J &= \{u,\bar{u},d,\bar{d},s,\bar{s},c,\bar{c},b,\bar{b}\}\,,~L = \{e^{+},e^{-},\mu^{+},\mu^{-},\tau^{+},\tau^{-},\nu_e,\nu_{\mu},\nu_{\tau}\}\,, \\
B &= \{W^+,W^-,Z,\gamma,h^0\}\,,
~T = \{t,\bar{t}\}\,, 
~H = \{h^0\}\,.
\label{bglabels}
\end{aligned}
\end{equation}
In general, events with gauge bosons and the SM Higgs boson in the final state are grouped into a single ``boson" (B) category. Thus, for example, the data set ``Bjj-vbf" represents production via vector boson fusion of a gauge boson or a Higgs boson with at least two additional light-quark jets.  The standard model background is displayed for two kinematic variables  $M_{\text{eff}}$(incl.) and $E_T^{\text{miss}}$ in Fig.~\ref{fig:bgs}.

In the analysis, production of all allowed sparticles in the final state for a given model point is carried out  using
 \code{MADGRAPH} at a center-of-mass energy of 14 TeV.
 The choice of 14 TeV rather than the current LHC 13 TeV-run energy is dictated by the SNOWMASS backgrounds used in this analysis which have been generated at 14 TeV. The cross-sections for all the model points of Table~\ref{tab1} are dominated by the production of $\tilde{\chi}^0_2\tilde{\chi}^{\pm}_1$ and $\tilde{\chi}^+_1\tilde{\chi}^-_1$ pairs. A non-negligible production of $\tilde{\chi}^0_1\tilde{\chi}^{\pm}_1$ pair is seen in most of the model points, while gluino production is greatly suppressed. A list of the different production cross-sections is shown in Table~\ref{tab3}. 
The myriad of the subsequent decay topologies following the production of the SUSY particles are the target of the signal region definitions and event selection chosen to enhance the signal-to-background ratio. The first signal region is based on jets and missing transverse energy with zero leptons in the final state. As Tables~\ref{tab4} and~\ref{tab5} indicate, the decays of the heavy neutralino and the chargino into jets and missing energy have the largest branching ratios among the other decays. The second set of signal regions is designed to search for leptons, jets and missing transverse energy in the final state. One of the signal regions in this set looks for a single lepton and the other for two leptons. The latter comes in two categories based on the flavor and charge of the leptons, namely, a pair of leptons having same flavor and opposite sign for the first category while the second looks for a pair with different flavor and opposite sign.   

Even though the signal regions used here are inspired by those from ATLAS and CMS analyses, the cuts and event selection for our analysis are {fully}  optimized to capture final states for the chargino coannihilation scenario whereby the mass gap between the chargino (and second neutralino) and the LSP is small and thus results in soft final states. 
   
\section{Event selection and results}\label{sec:signal}

Based on the character of the final states, event selection proceeds by employing a set of discriminating variables on the signal and SM background. Those variables have been optimized to give the best signal-to-background ratio by imposing tighter requirements on those variables so they are more sensitive to smaller mass gaps between the chargino and the LSP. The signal regions (SR) used in this analysis belong to three main categories: the zero lepton channel, the single {lepton} and the two lepton channels along with jets, namely, 0$\ell nj$, 1$\ell nj$ and 2$\ell nj$, where $n$ represents the minimum number of jets in the final state. The variety of SRs used have been inspired by analyses done by ATLAS and CMS~\cite{ATLAS:2017cjl, ATLAS:2017qih, ATLAS:2017uun, CMS:2015eoa} {and have been improved for the 
parameter sets of this work.}    

\subsection{Zero lepton channel}\label{zerolep}
The first signal region we investigate is the one that targets jets and missing transverse energy in the final state with a veto on all leptons (electrons and muons). This signal region, SR-0$\ell nj$, comprises of two signal regions based on the minimum number of jets in the final state, namely $0\ell 2j$ and $0\ell 4j$, with a minimum of 2 and 4 jets, respectively. The leading jets are required to have $p_T > 40$ GeV and all sub-leading jets have $p_T > 20$ GeV. A pre-cut of 100 GeV on the missing transverse energy, $E_{T}^{\text{miss}}$, is applied to the backgrounds and the signal. 
Table~\ref{tab6} shows the set of kinematic variables used in this signal region. Cuts are applied on the transverse momenta of jets $j_1$, $j_2$ and $j_4$ with an upper bound as high as 100 GeV for the leading jet. Jets are produced following the decay of a chargino or second neutralino to the LSP and with the available mass gap between the parent and daughter particles of up to 28 GeV, the extra kick in energy is evidently coming from ISR and FSR. We have two requirements on the azimuthal angle between any jet and the missing transverse energy, one is between the leading jet and the missing transverse energy, $\Delta\phi(\text{jet}_1,E_{T}^{\text{miss}})$, and the other is the smallest azimuthal separation between the same two objects, min[$\Delta\phi(\text{jet$_{1-2}$},E_{T}^{\text{miss}})$]. The latter is a good discriminator since background events from multi-jet processes tend to have a small value {for}  this variable. The kinematic variable $m_{\rm T2}$~\cite{Lester:1999tx, Barr:2003rg, Lester:2014yga} is used on the $0\ell2j$ signal region, which is defined as    
\begin{equation}
    m_{\rm T2}=\min\left[\max\left(m_{\rm T}(\mathbf{p}_{\rm T}(j_1),\mathbf{q}_{\rm T}),
    m_{\rm T}(\mathbf{p}_{\rm T}(j_2),\,\mathbf{p}_{\rm T}^{\text{miss}}-
    \mathbf{q}_{\rm T})\right)\right],
    \label{mt2}
\end{equation}
where $\mathbf{q}_{\rm T}$ is an arbitrary vector chosen to find the appropriate minimum and $m_{\rm T}$ is the transverse mass given by 
\begin{equation}
    m_{\rm T}(\mathbf{p}_{\rm T1},\mathbf{p}_{\rm T2})=
    \sqrt{2(p_{\rm T1}\,p_{\rm T2}-\mathbf{p}_{\rm T1}\cdot\mathbf{p}_{\rm T2})}.
\end{equation}

The minimum of the transverse masses of the jets and missing transverse energy, \\
$m^{\text{min}}_T(\text{jet$_{1-2}$},E_{T}^{\text{miss}})$, proves to be a good discriminator for the $0\ell 4j$ SR which reduces W+jets and $t\bar{t}$ background events, along with the effective mass, $m_{\text{eff}}$ defined as 
\begin{equation}
m_{\text{eff}} = \sum_{i\leq 4} (p_T^{\text{jets}})_i+E_{T}^{\text{miss}}. 
\end{equation}
The kinematic variable $m_{jj}$ is the invariant mass of the two leading jets and is used in the $0\ell 2j$ SR. Finally the variable $E_{T}^{\text{miss}}/\sqrt{H_T}$ uses $H_T$ defined as the scalar sum of the transverse momenta of all jets with a lower bound of 100 GeV being set on this variable for the SR $0\ell2j$ and 110 GeV for SR $0\ell4j$. All the selection criteria appearing in table~\ref{tab6} have been optimized to give the best minimum integrated luminosity for a $5\sigma$ discovery. The different SRs labelled A, B and C correspond to a variation of the kinematic variables $E_{T}^{\text{miss}}/\sqrt{H_T}$ and $H_T$ for $0\ell 2j$ and $H_T$ for $0\ell 4j$.   

The results obtained for the zero lepton channel with a minimum of 2 and 4 jets are shown in Table~\ref{tab7}, where entries with three dots indicate that the required integrated luminosity exceeds $3000\ifb$ which is the maximum value expected to be reached by the high luminosity LHC (HL-LHC). The distribution of integrated luminosities for the SR $0\ell2j$ ranges from 83$\ifb$ for point (e) in $0\ell 2j$-A to $3000\ifb$ for point (h) in $0\ell 2j$-C, while for $0\ell4j$ we have a low of $63\ifb$ for point (g) in $0\ell4j$-A and a high of $2960\ifb$ for point (j) also in $0\ell4j$-A. Out of the ten points, six of them are visible in all variations of both signal regions. \\
Figs.~\ref{fig2}-\ref{fig4} show some distributions in select kinematic variables used in SRs $0\ell2j$ and $0\ell4j$. In Fig.~\ref{fig2} distribution in the azimuthal separation between the leading jet and the missing energy, $\Delta\phi(\text{jet}_1,E_{T}^{\text{miss}})$, and distribution in the dijet invariant mass, $m_{jj}$, are plotted for point (a) in the SR $0\ell2j$-A at $86\ifb$. In $\Delta\phi(\text{jet}_1,E_{T}^{\text{miss}})$ the signal is above the background for larger values of this variable. Applying cuts at higher values of $\Delta\phi(\text{jet}_1,E_{T}^{\text{miss}})$ minimizes any misidentification of missing transverse momentum with jets. The distribution in $m_{jj}$ shows an excess of the signal above background for smaller values of this variable which is an indication of soft final states. This can also be seen in the distribution of the transverse momentum of the leading jet, $p_T(j_1)$, depicted in the left panel of Fig.~\ref{fig3}. As noted before, the extra kick is due to contributions from ISR and FSR jets. The right panel shows the distribution of the variable $E_{T}^{\text{miss}}/\sqrt{H_T}$ for point (e) at $83\ifb$ in the same SR. In Fig.~\ref{fig4} we exhibit three distributions in the variables $E_{T}^{\text{miss}}$, $m^{\text{min}}_T$ and $H_T$ for model point (b) at $1290\ifb$ in $0\ell4j$-C. The excess of the signal over the background is not as pronounced as in the examples before which is why this point requires a higher integrated luminosity for discovery in this particular SR.    

\subsection{Single lepton channel}\label{single_lepton}
Another SUSY signature to be discussed is the presence of a single isolated lepton in the final state coming from the decay of a chargino, along with jets and missing transverse energy. The signal region used here is labeled $1\ell2j$, where $2j$ indicates a minimum of two jets in the final state. Events are selected based on one tight electron or muon with $|\eta|<1.4$ for electrons and $|\eta|<1.2$ for muons. The azimuthal angle between the emitted lepton momentum and the missing energy is taken as $\Delta\phi(\vec{\ell},\vec{p}^{\text{miss}}_T)>1.5$ radians for the electrons and $\Delta\phi(\vec{\ell},\vec{p}^{\text{miss}}_T)<1.0$ radians for the muons. As for jets, it is required that the leading and 
the sub-leading jets both have $p_T > 20$ GeV. The other selection criteria for this SR are listed in Table~\ref{tab8}, where a 100 GeV pre-cut is also applied on $E_{T}^{\text{miss}}$. Here $m^{\ell}_T$ is the transverse mass of the lepton and $\vec{p}^{\text{miss}}_T$ and $m_{\text{eff}}$ is given by
\begin{equation} 
m_{\text{eff}} = p^{\ell}_T+\sum_{i\leq 2} (p_T^{\text{jets}})_i+E_{T}^{\text{miss}}.
\end{equation} 
The SRs A, B and C correspond to a variation of the transverse momentum of the leading jet. 
The minimum integrated luminosity for discovery is shown in Table~\ref{tab9}. It is clear that this SR is less successful than the previous one, with only 6 points being visible. The integrated luminosity ranges from $80\ifb$ for point (a) in $1\ell2j$-C to $2780\ifb$ for point (g) in $1\ell 2j$-B and C. In Fig.~\ref{fig5} we exhibit the distributions in the leptonic transverse mass, $m^{\ell}_T$ and the leading lepton transverse momentum, $p^{\ell}_T$ for model point (c) in SR $1\ell2j$-A at $2230\ifb$. Also here, the excess is over small values of those variables which explains the tight cuts applied in this SR.  

\subsection{Two lepton channel}

The last signal region we investigate is the presence of two leptons in the final state coming from the decay of the electroweakinos along with at least one jet. Events containing two leptons are selected such that the leading and 
the
sub-leading lepton transverse momenta must be $p^{\ell}_T >15$ GeV and 10 GeV, respectively. A veto on b-tagged jets is applied to reduce $t\bar{t}$ background events. This signal region, $2\ell1j$, contains two categories of SRs, one which looks for two leptons with same flavor and opposite sign (SFOS) and the other targets two leptons with different flavor and opposite sign (DFOS). For short, we label them as SF and DF. The kinematic variables used and the corresponding cuts are shown in Table~\ref{tab10}, where $\Delta R_{\ell\ell}$ and $m_{\ell\ell}$ represent, respectively, the separation between two SF or DF leptons and the invariant mass of those leptons. The cut on $m_{\ell\ell}$ ensures that background events corresponding to leptons coming from the decay of a $Z$ boson are reduced. The three signal regions A, B and C for each category correspond to the variation of the transverse mass $m_{\rm eff}$ defined by 
\begin{equation}
m_{\text{eff}} = \sum_{i\leq 2}(p^{\ell}_T)_i+ p_T(j_1)+E_{T}^{\text{miss}}.
\end{equation}
In order to reduce possible multi-jet backgrounds we use the variable $E_{T}^{\text{miss}}/H_T$ which is crucial in this SR. A series of optimizations have been carried out on this variable in order to reduce as much of the background 
as possible 
and retain as much of the signal as possible. Such optimization procedures are found useful in exploring
  atypical regions of the parameter space which could otherwise be missed (see, e.g.,~\cite{Kaufman:2015nda,Nath:2016kfp,Aboubrahim:2017aen,Chatterjee:2012qt}).
The resulting integrated luminosities for a 5$\sigma$ discovery are listed in Table~\ref{tab11} for the 10 benchmark points. The signal region $2\ell1j$-DF has a poor performance compared to $2\ell1j$-SF and thus has been eliminated. An integrated luminosity as low as $93\ifb$ is obtained for point (b) in $2\ell 1j$-SF-C and a high of $2800\ifb$ for point (j) also in $2\ell 1j$-SF-C. We exhibit in Fig.~\ref{fig6} the distributions in the dilepton invariant mass, $m_{\ell\ell}$, and the effective mass $m_{\rm eff}$. In the left panel, $m_{\ell\ell}$ is shown for model point (c) in $2\ell1j$-SF-C at $1590\ifb$. One can notice a major dip in the background events for $m_{\ell\ell}<10$ GeV and so cutting on this variable greatly improves the signal-to-background ratio. Another tight cut is applied on $m_{\rm eff}$ whose distribution is shown on the right panel for model point (d) at $334\ifb$. The signal is above background over a very narrow region which is again indicative of soft final states.

\subsection{Combined signal region results}

We combine now the results obtained thus far from the different signal regions used for analyzing the discovery potential of the supersymmetric models at the LHC. Table~\ref{tab12} shows the combined results with the leading and the sub-leading SRs and the corresponding integrated luminosities for a $5\sigma$ discovery.
  By the end of the LHC run II, ATLAS and CMS are expected to collect around $100\ifb$ of data each. 
  From  Table~\ref{tab12} we find  that the  parameter points, (a), (b), (e) and (g), would be within reach by the end of run II.  Further, all of the remaining parameter points of  Table~\ref{tab12}   will be
  discoverable in the high luminosity era of the LHC (HL-LHC) which  is expected to reach its optimal integrated luminosity of $3000\ifb$ at $\sqrt s=14$ TeV.

\section{The Gravitino Decay Constraints}
It is known that stable gravitinos produced in the early universe could overclose the universe if the gravitino
mass exceeds 1 keV~\cite{Pagels:1981ke}. Unstable gravitinos also produce cosmological constraints.
Since the gravitinos couple with the standard model fields gravitationally, they are long-lived and
 their decays could upset the BBN if they occur during or after the BBN time, i.e., $(1-10^2)$s. 
Of course the primordial gravitinos are all inflated away
during inflation but they can be regenerated in the reheating period after inflation.  So  we need to check
 the lifetime of the gravitinos for the benchmarks of Table~\ref{tab1}. A gravitino has many decay final states to the MSSM states which include the dominant two body decays 
 \begin{align}
 \tilde{G}\rightarrow \tilde{g}g, ~ & \tilde{\chi}_1^{\pm}W^{\mp}, ~\tilde{\chi}_1^0\gamma,
~\tilde{\chi}_1^0 Z.
\label{channels} 
 \end{align}
 In a general set up of  supergravity models, there is no direct equality of the gravitino mass and the scalar
 masses since the  scalar masses  could in general be nonuniversal depending on the form of the
 Kahler potential. However, here we make the simple assumption of the universality of the scalar masses
 at the GUT scale and the equality of the gravitino and the scalar mass, i.e., $m_{\tilde{G}}=m_0$.
  We point out, however, that scalar masses at the electroweak scale can differ significantly from 
 their values at the GUT scale  as a result of renormalization group evolution.  Thus from Table \ref{tab2} we see that  the masses of  the scalars  below the GUT scale are typically smaller than $m_0$ in the mass range
 investigated in Table \ref{tab1}.}
 In Table~\ref{tab13} we exhibit  the branching ratios of the leading gravitino decay channels of Eq.~(\ref{channels})
  along with the total decay width and the lifetime of the gravitino for the benchmarks of Table~\ref{tab1},
  where we have used the code \code{GravitinoPack}~\cite{Eberl:2013npa,Eberl:2015dia}. 
     Table~\ref{tab13} shows that for the benchmark of Table~\ref{tab1} the gravitino decays before the BBN time and thus BBN is not disturbed.

  However, there is still one further constraint from an unstable gravitino. Thus, although the gravitinos decay 
  before BBN, their decays produce neutralinos
 and if there is an overproduction of the gravitinos in the post inflationary period, their decay could generate 
  a neutrlalino relic density in excess of what is observed. Thus the relic density of neutralinos produced in the
  decay of the gravitinos acts as a constraint on the model.  
  So now we have the result that the total relic density of neutralinos is 
  \begin{align}
  \Omega_{\tilde{\chi}^0_1} =  \Omega^{\rm th}_{\tilde{\chi}^0_1} +  \Omega^{\G}_{\tilde{\chi}^0_1},  
  \label{total-omega}
  \end{align}
where the first term on the right hand side is from the conventional thermal production of neutralinos 
after freeze out and the second term is from the non-thermal contribution arising from the 
decay of the gravitino. Under the assumption that each gravitino decay results in just one neutralino we have
  \begin{align}
  \Omega^{\rm \G}_{\tilde{\chi}^0_1} =   \frac{m_{\tilde{\chi}_1^0}}{m_{\tilde G}}  \Omega_{\tilde G}.   
  \label{G-omega}
  \end{align}
Thus a computation of $\Omega^{\G}_{\tilde{\chi}^0_1}$ requires a computation of  $\Omega_{\tilde G}$
which depends on particulars of inflation  and specifically on the reheat temperature. 
Thus after the end of inflation, the inflaton field $\phi$  begins to  execute oscillations around the potential minimum.
 In a simplified treatment one makes the approximation that the coherent energy  of the inflaton is converted 
 instantaneously  into radiation energy at a time when the Hubble parameter $H\sim \Gamma_\phi$, 
 where $\Gamma_\phi$ is the decay width of the inflaton field $\phi$~\cite{Kolb:1990vq}.
 Thus one has the relation
\begin{align}
\rho_R= \rho_\phi|_{H= \Gamma_\phi},
\label{equality}
\end{align}
where $\rho_\phi$ is the energy density which on using the Friedmann equations in an FRW 
 universe with zero curvature 
  is given by 
\begin{align}
  \rho_\phi= \frac{3}{8\pi G_N} H^2\,, 
  \label{rho1}
\end{align}
where $G_N$ is Newton's constant.
Further, in Eq. (\ref{equality}) 
  $\rho_R$ is the radiation density which at the reheat temperature $T=T_R$ is given by
 \begin{align}
 \rho_R= \frac{\pi^2}{30} g_{*} T_{R}^4 \, ,
 \label{rho2}
 \end{align} 
  where  $g_*$ is the number of degrees of freedom at the reheat temperature $T_R$
  which for MSSM is $g_*=228.75$.
 Defining $M_{\rm Pl}= (\sqrt{8\pi G_N})^{-1/2}$ where  $M_{\rm Pl}$ is the reduced
Planck constant $M_{\rm Pl}\simeq 2.4\times 10^{18}$ GeV in Eq. \ref{rho1} and using Eq. \ref{equality}
one gets an expression for the reheat temperature 
 \begin{align}
 T_{R}=  \left(\frac{90}{\pi^2 g_*}\right)^{1/4} \sqrt{\Gamma_\phi \mP}\,.
  \end{align}
The above equation shows that the reheating depends on the details of the inflation model and specifically
through the decay width of the inflaton.  However, here we will
not go into the specifics of inflation models, of which there are many, but rather use the reheat temperature 
as our starting point which controls the thermal production of the gravitinos.  Of course the gravitinos can also be
produced by the decays of the inflaton, but again the branching ratio of the inflaton into the gravitino
is model dependent. For that reason we will focus on the thermal production of the gravitinos.

  The thermal production of gravitinos has been discussed in a variety of papers. A brief list of these  include~\cite{Ellis:1984eq,Giudice:1999am,Bolz:2000fu,Kohri:2005ru,Allahverdi:2005rh,Rychkov:2007uq,Pradler:2006qh,Pradler:2006hh,Arya:2016fnf}.
   In supersymmetric QCD the processes that produce the gravitino include  
 \begin{align}
 g\tilde g \to g\tilde G, ~g\tilde q \to q\tilde G, ~ q\bar q \to \tilde g \tilde G,\cdots 
 \end{align}
  In addition there are annihilation processes such as 
  $\tilde G \tilde G\to f\bar f,  \tilde g \tilde g$. However, we will ignore these back reactions since the gravitinos decouple at a  temperature 
    $\sim 10^{14}$ GeV, and thus they are decoupled from the thermal bath at the reheat temperatures 
   we consider below 
   which are significantly lower than the gravitino decoupling temperature.  It is found that the gravitino production cross section is proportional to  the sum of two terms, one
   from     the 
  production of $\pm 3/2$ helicity states and the other from  the production of $\pm 1/2$ helicity states. 
   Thus the Boltzmann equation governing the thermal production of gravitinos after reheating is given by 
 \begin{align}
    \frac{dn_{\tilde G}}{dt} + 3 H n_{\tilde G} = a_{\tilde G},
    \label{bolt1}
   \end{align}  
   where~\cite{Pradler:2006qh}   
 \begin{align}
 a_{\tilde G} =
           \frac{3\zeta(3)T^6}{16\pi^3M_{\rm Pl}^2} 
   \sum_{i=1}^{3}  c_i g_i^2\left(1+\frac{m_i^2}{3 m_{\tilde G}^2}\right)
        \ln\left(\frac{k_i}{g_i}\right).    
        \label{bolt2}
\end{align}
 Here $m_i$ ($i=1,2,3$) are the gaugino masses for the gauge groups $U(1)_Y$, $SU(2)_L$ and $SU(3)_C$
 and $g_i$ are the  corresponding gauge coupling constants where $m_i$ and $g_i$ are 
 evaluated at temperature $T$. 
 Further, $c_i=(11, 27,72)$ and 
 $k_i=(1.266, 1.312, 1.271)$. 
  We note that Eq.~(\ref{bolt2}) contains the factor 
  \begin{align}
   \frac{1}{M_{\rm Pl}^2}\left(1+ \frac{m_i^2}{3 m_{\tilde G}^2}\right).
  \end{align}
  The significance of this factor is the following: 
  the first term in the brace arises from the  production of the $\pm 3/2$ helicity states  
 of the gravitino while the second term in the brace arises from the production 
  of  $\pm 1/2$ helicity components.
  Note that the term that arises from  $\pm 3/2$ helicities is independent of  $m_i$ and $m_{\tilde G}$ while
  the term that arises from $\pm 1/2$ helicities is dependent on both $m_i$ and $m_{\tilde G}$. 

\begin{figure}[t]
\centering
\includegraphics[width=.6\textwidth]{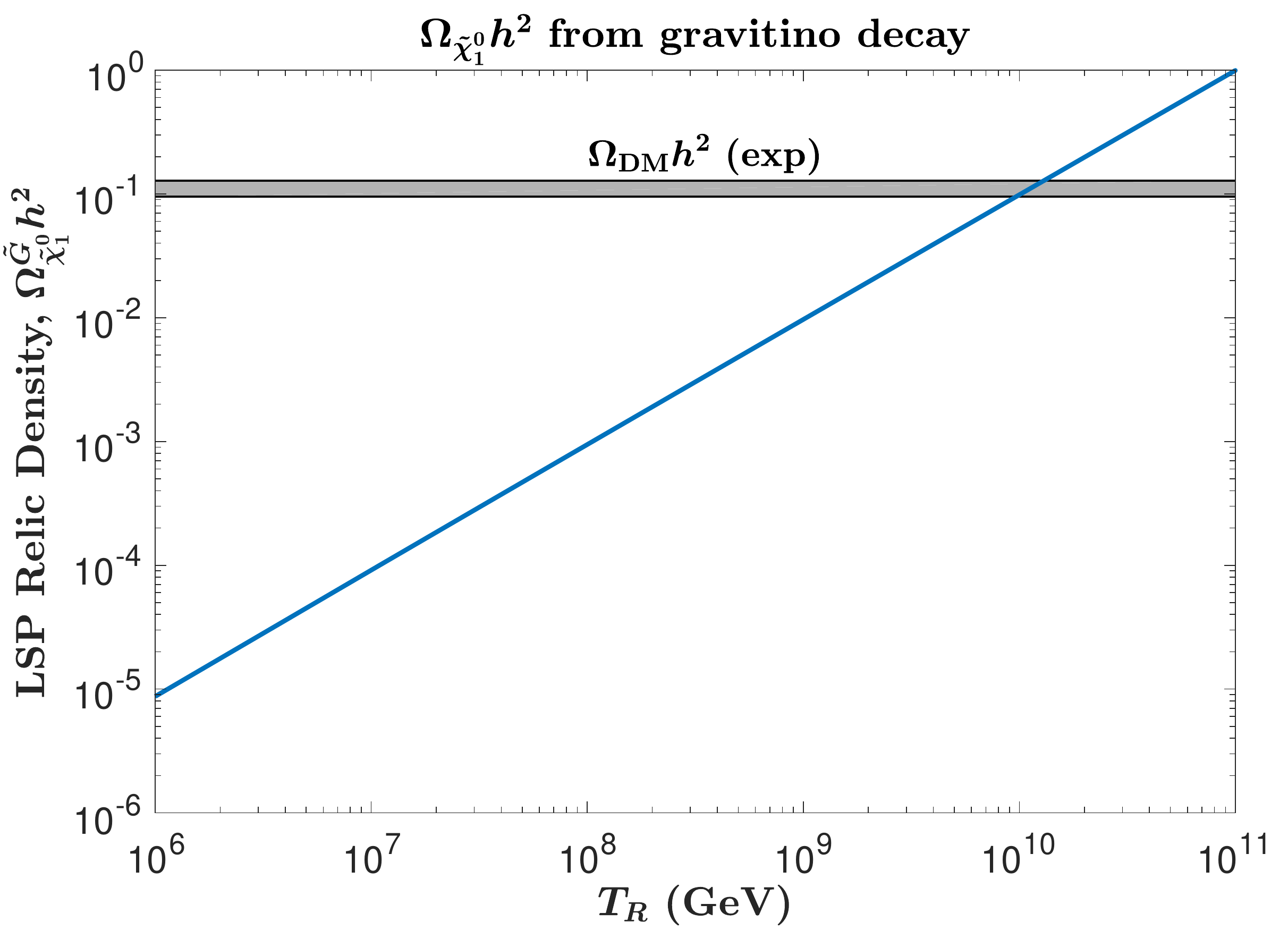}
\caption{Neutralino relic density arising from the decay of thermally produced gravitinos vs the reheat temperature 
using the model points of Table \ref{tab1}. All the model points lie on the thin blue line.
The squeezing of the model points on to the thin blue line arises 
 because for $m_i^2/m^2_{\tilde G}<<1$ the gravitino  production is dominated by $\pm 3/2$ helicity states 
while the production of $\pm 1/2$ helicity  states is suppressed by a factor  $(m_i^2/m^2_{\tilde G})$
(see Eq.~\ref{relic-G}). The grey patch correspond to $\Omega_{\rm DM}h^2$ (exp), the experimental limits on dark matter relics from the WMAP and PLANCK experiments.}
\label{fig16}
\end{figure}

Eq.~\ref{bolt1} can be solved  analytically under the assumption of conservation of entropy 
per comoving volume~\cite{Bolz:2000fu}.
Here we use Eq.(3) of  \cite{Pradler:2006qh} in  Eq.  \ref{G-omega}
to obtain the neutralino relic density arising from the decay of the gravitino so that 
\begin{align}
        \Omega^{\G}_{\chi_1^0}h^2
        &=
        \sum_{i=1}^{3}
        \omega_i\, g_i^2 
        \left(1+\frac{m_i^2}{3m_{\tilde G}^2}\right)
        \ln\left(\frac{k_i}{g_i}\right)
        \left(\frac{m_{\tilde{\chi}_1^0}}{100 \GeV}\right)
        \left(\frac{T_R}{10^{10} \GeV}\right).
\label{relic-G}
\end{align}
 Here $\omega_i(i=1,2,3)= (0.018, 0.044, 0.177)$~\cite{Pradler:2006qh} and $m_i$ and $g_i$ are evaluated 
 at temperature $T_R$. They can be obtained from their GUT values by using the relations
 \begin{align}
\frac{m_{i}(T_R)}{m_{i}(M_G)}&=\frac{g^2_{i}(T_R)}{g^2_{i}(M_G)}\, ,\nonumber\\
\frac{1}{g^2_{i}(T_R)}&= \frac{1}{g^2_i(M_G)} + 
\frac{\beta_i^{(i)}}{8 \pi^2}\ln\left(\frac{M_G}{T_R}\right).
\end{align}
Here $M_G$ is the GUT scale, $g_i(T_R), m_i(T_R)$ are the gauge couplings  and the gaugino masses
at $T_R$, and  $g_i(M_G), m_i(M_G)$ are their GUT values, $\beta_i^{(1)}$ are the one loop evolution coefficients given by
$\beta_i^{(1)}(i=1,2,3)= (11, 1, -3)$. 
  Numerical result of the relic density of neutralinos  
 produced via decay of the gravitino vs the reheat temperature  $T_R$  is exhibited in Fig ~\ref{fig16}. 
 All the model points given in Table~\ref{tab1} lie on the thin blue line. The insensitivity of the neutralino relic density to the gravitino mass is easily understood from Eq.~\ref{relic-G} since the relic density becomes
 independent of the gravitino mass in the limit $m_i/m_{\tilde G} <<1$ which is the case for the model
 points of Table~\ref{tab1}. The analysis of Fig~\ref{fig16} shows that the neutralino relic density arising
 from gravitino decay is below the relic density given by WMAP and PLANCK up to reheat temperature
 of $10^{10}$ GeV and is a negligible fraction of the total
 for reheat temperatures below $10^{9}$ GeV.
 The deduction of the reheat temperature is rather model
 dependent since it involves the nature of the inflaton, its coupling to the standard model fields and the 
 possible modes of its decay, i.e., gauge, Yukawa or gravitational. Thus the analysis presented above is in
 terms of the reheat temperature rather than in terms of an underlying inflaton model.

\section{Dark matter in SUGRA with 50-100 TeV scalars}
The analysis presented in Table~\ref{tab12} gives us a set of models which are consistent with  the Higgs boson mass constraint and  the relic density consistent with the WMAP
 and PLANCK  experiments and would be discoverable at the LHC with an integrated luminosity  well below the 
 optimal integrated luminosity achievable at the LHC. 
 It is also of interest to investigate if some or all of these  models are discoverable in direct detection experiment.
  The direct detection of the neutralinos depends crucially on its gaugino-higgsino content.
 Thus the neutralino is a linear combination of four states
$\tilde \chi^0= \alpha \lambda^0 + \beta \lambda^3 + \gamma \tilde H_1+ \delta \tilde H_2$ where $\lambda^0, \lambda^3$
are the bino, wino  and $\tilde H_1, \tilde H_2$ are the higgsinos.  For the models of Table~\ref{tab1},
$|\beta|\leq 0.324,  |\gamma|\leq 0.003, |\delta|=0.000$.
One finds that the wino and the higgsino content of the models of Table~\ref{tab1} are small, and the neutralino
is essentially a bino. The fact that the neutralino is mostly a bino makes the neutralino-proton cross section relatively small. 
 In Table~\ref{tab14} we present the spin independent and spin-dependent neutralino-proton cross sections 
 for these models. The analysis of Table~\ref{tab14} shows that the spin-independent neutralino-proton cross section is $\mathcal{O}(10^{-48}$cm$^{-2})$,
 and  only three out of the ten benchmarks lie above the neutrino floor~\cite{Strigari:2009bq} 
 which is the  threshold for detectability (see Fig.~\ref{sigma}). Those points   would thus be out of reach  of the future dark matter experiments LUX-ZEPLIN~\cite{Schumann:2015wfa,Cushman:2013zza}.  However, as Table~\ref{tab12} shows they would be discoverable at the LHC.

\section{Conclusions}

Supersymmetry is desirable for a number of theoretical as well as phenomenological reasons.
Supergravity unification provides a framework  with a small number of parameters at a  high scale
 in terms of which the properties of low energy effective theory can be computed.  Supergravity unified models also accomplish
 radiative breaking of the electroweak symmetry which allows a determination of the sparticle mass spectrum with the given 
 high scale input and thus determines the weak SUSY scale.   
 The observation of the Higgs boson mass at $\sim 125$ GeV implies that the loop correction to the tree
level Higgs boson mass  is large which in turn implies that the scale of weak scale supersymmetry 
lies in the  several TeV region. This makes the search for supersymmetry more challenging than initially 
thought. For high scale models, there is another aspect which makes the observation of 
supersymmetry challenging. This concerns dark matter.   Thus for  high scale models one finds that often the parameter space that gives the desired Higgs boson mass gives a neutralino which is mostly a bino.  
For a bino type neutralino, one needs \co to achieve the appropriate relic density consistent with the  WMAP and PLANCK experiments. This means that there must be one or more sparticles close by to coannihilate with the neutralino. The relatively small  mass gap between the neutralino and the coannihilating particles implies that the final states in the  decay of the coannihilation  process would be soft and thus hard to detect.  

In this analysis we have investigated high scale models with scalars in the mass range 50-100 TeV  while the gauginos are relatively light. 
 Scalar masses in the assumed range are interesting as they alleviate a number of problems associated with 
 low weak SUSY scale.
 One such problem concerns the SUSY CP problem which leads to large EDMs for leptons and quarks 
 significantly above the existing experimental limits. Aside from fine tuning the CP phases to be 
 extremely small, the other options include mass suppression \cite{suppression} or the cancellation mechanism
 ~\cite{cancellation}. The models with  scalar masses in the 50-100 TeV naturally provide a large mass 
 suppression of the EDMs alleviating this problem in a significant way.  Another problem of low weak SUSY scale
 concerns rapid proton decay even with R-parity conservation due to baryon and lepton number violating dimension five operators. Again scalar masses in the 50-100 TeV range resolve this problem in a natural way~\cite{Liu:2013ula}.
 In the analysis presented here we created a number of benchmark models consistent with radiative electroweak symmetry  breaking, Higgs boson mass constraint and the relic density
 constraint on neutralino dark matter. We used an extensive set of signal regions and optimized them for the model points we discuss. 
In the analysis we found a number of  signatures with 0, 1 and 2 leptons, 2 and 4 jets, along with other kinematical constraints which
 allow a $5\sigma$ discovery for some of our  benchmarks  with an integrated luminosity of  100 fb$^{-1}$. All of the remaining benchmarks are found to be discoverable with an integrated luminosity of 1000 fb$^{-1}$ which is significantly
 below the optimum integrated luminosity that can be reached in the high luminosity era of the LHC. 
 
  We also investigated the influence of 50-100 TeV scalar masses on unification of gauge coupling constants. It is 
  found that unification of the gauge couplings using LEP data occurs with the same degree of accuracy as for 
  the case with weak scale supersymmetry in the TeV region. Further, we analyzed the decay of the gravitinos in this
  model and found that the gravitinos decay before the time scale  $(1-10^2)$s and do not upset the BBN. Further, we analyzed the thermal
  production of gravitinos and their contribution to the non-thermal relic density of the neutralinos.
  Here one finds that this contribution is negligible up to reheat temperature of 
  $10^{9}$ GeV.  
 We also analyzed the 
spin-independent neutralino-proton cross section. It is found that only few of those points have cross sections lying above the neutrino floor 
while others are below it making them difficult to detect  even with future dark matter experiments such as LUX,  ZEPLIN and XENON1T. 
Thus the latter set could only be discovered at the LHC. In summary, high scale models with scalar masses lying in the 50-100 TeV have the
possibility of being  discovered at the LHC and such models also have several redeeming properties as they alleviate some of the
problems encountered by low weak scale SUSY models. 

Finally we note the analysis above exhibits the remarkable effect of dark matter constraints through 
\co on limiting the parameter space of models and thus controlling the discovery potential of the LHC for 
supersymmetry.  One can thus expect some influence on the LHC anaylses if the nature of dark matter was 
not pure neutralino but was multi-component  (see e.g.,~\cite{Feldman:2010wy}).
 One such possibility proposed recently is in the form of  an ultralight axion\cite{Marsh:2015xka}. 
  If this were the case the relic density arising from neutralino would decrease making the dark 
 matter constraint on the analysis more stringent. However, at this time there is no compelling evidence 
 for the multicomponent nature of dark matter.  

\textbf{Acknowledgments: }
Correspondences with Howard Baer and Tom Cridge are acknowledged.  The analysis presented here was done
using the resources of the high-performance  Cluster353 at the Advanced Scientific Computing Initiative (ASCI) 
at Northeastern University.  This research was supported in part by the NSF Grant PHY-1620575.

\section{Tables} 

\begin{table}[H]
\begin{center}
\begin{tabulary}{0.85\textwidth}{l|CCCCCC}
\hline\hline\rule{0pt}{3ex}
Model & $m_0$ & $A_0$ & $m_1$ & $m_2$ & $m_3$ & $\tan\beta$ \\
\hline\rule{0pt}{3ex}  
\!\!(a)  & 70760  & 141410 & 544 & 481& 983 & 45 \\
(b)  & 77710  & 155593 & 503 & 426 & 1645 & 11 \\
(c)  & 92390 & 183892 & 557 & 474 & 1441 & 18 \\
(d)  & 82900  & 165862 & 539 & 466 & 1275 & 6 \\
(e)  & 63057  & 126110 & 504 & 414 & 1472 & 28 \\
(f)  & 67248 & 134496 & 543 & 446 & 1482 & 30 \\
(g) & 54981 & 109990 & 521 & 419 & 1388 & 34 \\
(h)  & 86618  & 172526 & 610 & 497 & 1369 & 23 \\
(i)  & 58619 & 117055 & 550 & 425 & 1204 & 25 \\
(j)  & 74199  & 148386 & 620 & 487 & 1000 & 27 \\
\hline
\end{tabulary}\end{center}
\caption{Input parameters for benchmarks for high weak scale supergravity models with $m_0$ in the range
$50-100$ TeV range. All masses are in GeV.}
\label{tab1}
\end{table}

\begin{table}[H]
\begin{center}
\begin{tabulary}{0.85\textwidth}{l|CCCCCCCCC}
\hline\hline\rule{0pt}{3ex}
Model  & $h^0$ & $\tilde\mu$ & $\tilde\chi_1^0$ & $\tilde\chi_1^\pm$ & $\tilde t$ & $\tilde g$ & $m_{\tilde q}$ & $m_{\tilde f}$ & $\Omega^{\rm th}_{\tilde{\chi}^0_1} h^2$ \\
\hline\rule{0pt}{3ex} 
\!\!(a) & 124.9 & 21006 & 114.1 & 134.2 & 36219 & 2149 & 70362 & 63300 & 0.103 \\
(b) & 125.5 & 30025 & 129.2 & 144.5 & 37578 & 3799 & 77314 & 73600 & 0.108 \\
(c) & 124.4 & 31386 & 130.0 & 145.9 & 46109 & 3336 & 91885 & 87000 & 0.126 \\
(d) & 124.8 & 34655 & 136.1 & 152.5 & 40236 & 2922 & 82456 & 78700 & 0.126 \\
(e) & 124.9 & 23457 & 136.2 & 156.0 & 30474 & 3403 & 62740 & 58500 & 0.113 \\
(f) & 123.6 & 24163  & 147.3 & 169.4 & 32873 & 3405 & 66909 & 62200 & 0.127 \\
(g) & 123.2 & 20755 & 147.7 & 170.6 & 26391 & 3211 & 54707 & 50400 & 0.115 \\
(h) & 126.7 & 27898 & 169.4 & 192.5 & 43771 & 3106 & 86200 & 81200 & 0.115 \\
(i) & 123.7 & 21964 & 170.9 & 195.3 & 28254 & 2795 & 58300 & 54700 & 0.114 \\
(j) & 124.3 & 24222 & 187.1 & 214.9 & 37276 & 2252 & 73800 & 69100 & 0.085 \\

\hline
\end{tabulary}\end{center}
\caption{The Higgs boson ($h^0$) mass, some relevant sparticle masses, and the relic density for the benchmarks of Table~\ref{tab1}. { Here $m_{\tilde{q}}$ stands for}  the mass of the heaviest squark and $m_{\tilde{f}}$
{for} the average sfermion mass. All masses are in GeV. }
\label{tab2}
\end{table}

\begin{table}[H]
\begin{center}
\begin{tabulary}{1.05\textwidth}{l|CCCCCC}
\hline\hline\rule{0pt}{3ex}
Model & full SUSY & $qq\rightarrow\tilde\chi_2^0\tilde\chi_1^\pm$ & $qq\rightarrow\tilde\chi_1^+\tilde\chi_1^-$  & $qq\rightarrow\tilde\chi_1^0\tilde\chi_1^{\pm}$ & $qq\rightarrow\tilde{g}\tilde{g}$ & $gg\rightarrow\tilde{g}\tilde{g}$ \\
\hline\rule{0pt}{3ex}
\!\!(a) & 9.45 & 5.82 & 3.08 & 0.55 & $2.4\times 10^{-4}$ & $1.7\times 10^{-4}$ \\
(b) & 7.16 & 4.79 & 2.37 & $3.7\times 10^{-5}$ & $1.9\times 10^{-7}$ & $1.3\times 10^{-8}$ \\
(c) & 6.92 & 4.63 & 2.29 & $1.3\times 10^{-4}$ & $1.3\times 10^{-6}$ & $2.1\times 10^{-7}$ \\
(d) & 5.91 & 3.96 & 1.95 & $8.7\times 10^{-6}$ & $7.9\times 10^{-6}$ & $2.3\times 10^{-6}$ \\
(e) & 5.50 & 3.47 & 1.80 & 0.22 & $9.8\times 10^{-7}$ & $1.4\times 10^{-7}$ \\
(f) & 4.10 & 2.57 & 1.34 & 0.19 & $9.7\times 10^{-7}$ & $1.4\times 10^{-7}$ \\
(g) & 4.00 & 2.48 & 1.30 & 0.22 & $2.2\times 10^{-6}$ & $4.4\times 10^{-7}$ \\
(h) & 2.58 & 1.62 & 0.84 & 0.12 & $3.6\times 10^{-6}$ & $7.8\times 10^{-7}$ \\
(i) & 2.46 & 1.52 & 0.80 & 0.14 & $1.4\times 10^{-5}$ & $4.6\times 10^{-6}$ \\
(j) & 1.74 & 1.03 & 0.56 & 0.15 & $1.5\times 10^{-4}$ & $9.6\times 10^{-5}$ \\
\hline
\end{tabulary}\end{center}
\caption{SUSY production cross sections, in pico-barns, for benchmarks of Table~\ref{tab1} {where ``full SUSY'' stands for the total production cross section including all the sparticle final states in the production.}}
\label{tab3}
\end{table}  

\begin{table}[H]
\begin{center}
\begin{tabulary}{2.00\textwidth}{l|CCCCC}
\hline\hline\rule{0pt}{3ex}
Model & $\tilde\chi_2^0\to\tilde\chi_1^0 q\bar{q}$ & $\tilde\chi_2^0\to\tilde\chi_1^0\ell\bar{\ell}$  & $\tilde{g}\to\tilde\chi_1^0 q\bar{q}$ & $\tilde{g}\to\tilde\chi_2^0 q\bar{q}$ & $\tilde{g}\to\tilde\chi_1^{\pm} q_{i}\bar{q}_j$ \\
& $q\in\{u,d,c,s,b\}$ & $\ell\in\{e,\mu,\tau,\nu\}$ & \multicolumn{2}{c}{$q\in\{u,d,c,s,t,b\}$} & \\
\hline\rule{0pt}{3ex}
\!\!(a) & 0.73 & 0.27 & 0.18 & 0.27 & 0.55\\
(b) & 0.71 & 0.29 & 0.28 & 0.24 & 0.48 \\
(c) & 0.72 & 0.28 & 0.25 & 0.25 & 0.50 \\
(d) & 0.66 & 0.34 & 0.28 & 0.24 & 0.48 \\
(e) & 0.76 & 0.24 & 0.25 & 0.25 & 0.50\\
(f) & 0.76 & 0.24 & 0.24 & 0.25 & 0.51\\
(g) & 0.77 & 0.23 & 0.24 & 0.25 & 0.51\\
(h) & 0.74 & 0.26 & 0.24 & 0.25 & 0.51\\
(i) & 0.76 & 0.24 & 0.26 & 0.25 & 0.49\\
(j) & 0.75 & 0.25 & 0.23 & 0.25 & 0.52\\
\hline
\end{tabulary}\end{center}
\caption{Branching ratios for dominant decays of $\tilde\chi_2^0$ and $\tilde{g}$ for benchmarks of Table~\ref{tab1}
where  $q_i\bar{q}_j=\{(u\bar{d}), (d\bar{u}), (s\bar{c}), (c\bar{s}), (b\bar{t}), (t\bar{b})\}.$}
\label{tab4}
\end{table}

\begin{table}[H]
\begin{center}
\begin{tabulary}{2.00\textwidth}{l|CC}
\hline\hline\rule{0pt}{3ex}
Model & $\tilde{\chi}_1^{\pm}\to\tilde\chi_1^0 q_i\bar{q_j}$ & $\tilde{\chi}_1^{\pm}\to\tilde\chi_1^0 \ell^{\pm}\nu_{\ell}$\\
& $q\in\{u,d,c,s\}$ & $\ell\in\{e,\mu,\tau\}$\\
\hline\rule{0pt}{3ex}
\!\!(a) & 0.67 & 0.33 \\
(b) & 0.67 & 0.33 \\
(c) & 0.67 & 0.33 \\
(d) & 0.67 & 0.33 \\
(e) & 0.67 & 0.33 \\
(f) & 0.67 & 0.33 \\
(g) & 0.67 & 0.33 \\
(h) & 0.67 & 0.33 \\
(i) & 0.67 & 0.33 \\
(j) & 0.67 & 0.33 \\
\hline
\end{tabulary}\end{center}
\caption{Branching ratios for dominant decays of $\tilde\chi_1^\pm$ for benchmarks of Table~\ref{tab1}
where $q_i\bar{q}_j=\{(u\bar{d}), (c\bar{s}), (t\bar{b})\}.$}
\label{tab5}
\end{table}

\begin{table}[H]
	\begin{center}
	\begin{tabulary}{0.85\linewidth}{l|cccccc}
    \hline\hline
	Requirement & \multicolumn{6}{c}{0$\ell nj$} \\
    \hline
    & $0\ell 2j$-A & $0\ell 2j$-B & $0\ell 2j$-C & $0\ell 4j$-A & $0\ell 4j$-B & $0\ell 4j$-C \\
    \cline{2-7}
    $N$(jets)
    & $\geq 2$ & $\geq 2$ & $\geq 2$ & $\geq 4$ & $\geq 4$ & $\geq 4$ \\
    $p_T(j_1)\text{ (GeV)}<$
    & 100 & 100 & 100 & 100 & 100 & 100  \\
    $p_T(j_2)\text{ (GeV)}<$
    & 60 & 60 & 60 & 80 & 80 & 80  \\
    $p_T(j_4)\text{ (GeV)} <$
    &  &  &  & 50 & 50 & 50  \\
    $E_{T}^{\text{miss}} \text{(GeV)}<$
    & 250 & 250 & 250 & 400 & 400 & 400  \\
    $\Delta\phi(\text{jet}_1,E_{T}^{\text{miss}}) \text{(rad)}>$
    & 1.5 & 1.5 & 1.5 & 2.5 & 2.5 & 2.5 \\
    min[$\Delta\phi(\text{jet}_{1-2},E_{T}^{\text{miss}}$)] $\text{(rad)}<$
    & 2.5 & 2.5 & 2.5 &  &  & \\
    $m_{\rm T2} \text{(GeV)} >$
    & 100 & 100 & 100 & & &  \\
    $m_{\rm T2} \text{(GeV)} <$
    & 400 & 400 & 400 & & &  \\
    $m_{jj} \text{(GeV)} >$
    & 50 & 50 & 50 &  & &  \\
    $m_{jj} \text{(GeV)} <$
    & 700 & 700 & 700 &  & & \\
    $m^{\text{min}}_T(\text{jet}_{1-2},E^{\text{miss}}_T) \text{ (GeV)}<$
    &  &  &  & 120 & 120 & 120 \\
    $m_{\text{eff}} \text{(GeV)} >$
    &  &  &  & 250 & 250 & 250  \\
    $m_{\text{eff}} \text{(GeV)} <$
    &  &  &  & 350 & 350 & 350 \\
    $E_{T}^{\text{miss}}/\sqrt{H_T} \text{(GeV}^{1/2})>$
    & 1 & 1 & 1 &  &  &   \\
    $E_{T}^{\text{miss}}/\sqrt{H_T} \text{(GeV}^{1/2})<$
    & 15 & 15 & 13 &  &  &   \\
    $H_{T} \text{(GeV)}<$
    & 115 & 120 & 120 & 155 & 160 & 165 \\
    \hline
     \end{tabulary}\end{center}
	\caption{The selection criteria $(0\ell nj)$ used for the signal regions {implies that the signal consists of  zero leptons (veto on electrons and muons) and $n$ jets where
	$n$ is  a minimum of 2 or 4 jets in the final state. The blank spaces indicate that the kinematical variable is either not applicable to the corresponding SR or has not been used.} }
	\label{tab6}
\end{table}

\begin{table}[H]
	\centering
	\begin{tabulary}{\linewidth}{l|cccccccc}
    \hline\hline
    & \multicolumn{6}{c}{$\mathcal{L}$ for $5\sigma$ discovery in 0$\ell nj$}  \\
	\hline
	Model & $0\ell 2j$-A & $0\ell 2j$-B & $0\ell 2j$-C & $0\ell 4j$-A & $0\ell 4j$-B & $0\ell 4j$-C \\
	\hline
  (a)  & 86 & 431 & 492 & 100 & 422 & 417 \\
  (b)  & 150 & 920 & 990 & 175 & 734 & 1290 \\
  (c)  & 114 & 1060 & 1140 & 749 & ... & ...\\
  (d)  & 174 & 864 & 864 &  &  & \\
  (e) & 83 & 848 & 894 & 74 & 199 & 548 \\
  (f) & 173 & 1250 & 1450 & 534 & 560 & 1420 \\
  (g) & 196 & 1250 & 1250 & 63 & 147 & 460 \\
  (h)  & 558 & 2870 & 3000 & 337 & 905 & 1830 \\
  (i) & 1120 & ... & ...  & 1480 & 1560 & 2010 \\
  (j)  & 771 & ... & ... & 2960 & ... & ... \\
	\hline
	\end{tabulary}
	\caption{Analysis of the discovery potential for supersymmetry for the parameter {set}  of Table~\ref{tab1}, using the selection criteria of Table~\ref{tab6}, where the minimum integrated luminosity needed for $5\sigma$ discovery is given in fb$^{-1}$. Here and in the tables following ... indicates that the minimum integrated luminosity needed for 5$\sigma$ discovery exceeds 3000 fb$^{-1}$. Blank spaces mean that zero events passed the cuts.}
\label{tab7}
\end{table}

\begin{table}[H]
	\begin{center}
	\begin{tabulary}{0.85\linewidth}{l|c|c|c}
    \hline\hline
	Requirement & \multicolumn{3}{c}{1$\ell 2j$} \\
    \hline
    & $1\ell 2j$-A & $1\ell 2j$-B & $1\ell 2j$-C \\
    \cline{2-4}
    $N$(jets)
    & $\geq 2$ & $\geq 2$ & $\geq 2$ \\
    $p_T(j_1)\text{ (GeV)}<$
    & 60 & 70 & 80  \\
    $p_T(j_2)\text{ (GeV)}<$
    & 50 & 50 & 50  \\
    $\text{Leading} \, p^{\ell}_T \text{ (GeV)}>$
    & 10 & 10 & 10  \\
    $\text{Leading} \, p^{\ell}_T \text{ (GeV)}<$
    & 40 & 40 & 40 \\
    $m^{\ell}_T \text{ (GeV)}<$
    & 60 & 60 & 60  \\
    $m^{\text{min}}_T(\text{jet}_{1-2},E^{\text{miss}}_T) \text{ (GeV)}<$
    & 140 & 140 & 140 \\
    $m_{\text{eff}} \text{ (GeV)}>$
    & 180 & 180 & 180  \\
    $m_{\text{eff}} \text{ (GeV)}<$
    & 240 & 240 & 240  \\
    $E_{T}^{\text{miss}} \text{(GeV)}<$
    & 250 & 250 & 250  \\
    $\Delta\phi(\text{jet}_1,E_{T}^{\text{miss}}) \text{(rad)}>$
    & 2.5 & 2.5 & 2.5 \\
    $H_{T} \text{(GeV)}<$
    & 105 & 105 & 105  \\
    \hline
     \end{tabulary}\end{center}
	\caption{The selection criteria {$(1\ell 2j)$} used for the signal regions corresponding to a single lepton, missing transverse energy and a minimum of 2 jets in the final state.}
	\label{tab8}
\end{table}

\begin{table}[H]
	\centering
	\begin{tabulary}{\linewidth}{l|ccc}
    \hline\hline
    & \multicolumn{3}{c}{$\mathcal{L}$ for $5\sigma$ discovery in 1L2J}  \\
	\hline
	Model & $1\ell 2j$-A & $1\ell 2j$-B & $1\ell 2j$-C \\
	\hline
  (a)  & 1200 & 221 & 80 \\
  (b)  & 520 & 385 & 217  \\
  (c)  & 2230 & ... & 928 \\
  (f) & ... & ... & 2640 \\
  (g) & 1670 & 2780 & 2780  \\
  (h)  & ... & 1670 & 1070 \\
	\hline
	\end{tabulary}
	\caption{Analysis of the discovery potential for supersymmetry for the parameter space of Table~\ref{tab1}, using the selection criteria of Table~\ref{tab8}, where the minimum integrated luminosity needed for $5\sigma$ discovery is given in fb$^{-1}$. Points (d), (e), (i) and (j) are not displayed since the minimum integrated luminosity needed for their discovery exceeds 3000 fb$^{-1}$ for this signal region.}
\label{tab9}
\end{table}

\begin{table}[H]
	\begin{center}
	\begin{tabulary}{\linewidth}{lccccccc}
    \hline\hline
	 & \multicolumn{3}{c}{2$\ell 1j$-SF} & & \multicolumn{3}{c}{2$\ell 1j$-DF} \\
    \cline{2-4} \cline{6-8}
    Requirement & $2\ell1j$-SF-A & $2\ell1j$-SF-B & $2\ell1j$-SF-C & & $2\ell1j$-DF-A & $2\ell1j$-DF-B & $2\ell1j$-DF-C \\
    \hline
    $E^{\text{miss}}_T\text{ (GeV)}<$
    & 150 & 150 & 150 & & 150 & 150 & 150 \\
    $m^{\ell}_T \text{ (GeV)}<$
    & 80 & 80 & 80 & & 80 & 80 & 80 \\
    $\Delta\phi(j_1, E^{\text{miss}}_T)\text{ (rad)}>$
    & 2.7 & 2.7 & 2.7 & & 2.7 & 2.7 & 2.7 \\
    $\Delta R_{\ell\ell}\text{ (rad)}<$
    & 1.0 & 1.0 & 1.0 & & 3.0 & 3.0 & 3.0 \\
    $m_{\ell\ell}\text{ (GeV)}<$
    & 50 & 50 & 50 & & 40 & 40 & 40 \\
    $E^{\text{miss}}_T/H_{T} > $
    & 0.7 & 0.7 & 0.7 & & 0.7 & 0.7 & 0.7 \\
    $m_{\rm eff}\text{ (GeV)}>$
    & 160 & 160 & 160 & & 160 & 160 & 160 \\
    $m_{\rm eff}\text{ (GeV)}<$
    & 260 & 270 & 280 & & 260 & 270 & 280 \\
	\hline
	\end{tabulary}\end{center}
	\caption{The selection criteria used for the signal regions related to the 2 lepton signature. Here and in the tables following SF stands for 
	same flavor opposite sign lepton pair  and DF stands for different flavor opposite sign lepton pair.}
	\label{tab10}
\end{table}

\begin{table}[H]
	\begin{center}
	\begin{tabulary}{\linewidth}{l|ccc}
    \hline\hline
	 & \multicolumn{3}{c}{$\mathcal{L}$ for $5\sigma$ discovery in 2$\ell 1j$-SF} \\
    \cline{2-4} 
    Model & $2\ell1j$-SF-A & $2\ell1j$-SF-B & $2\ell1j$-SF-C  \\
    \hline
    (a) & 131 & 167 & 214  \\
    (b) & 228 & 291 & 93 \\
    (c) & 975 & 1250 & 1590  \\
    (d) & 334 & 427 & 243 \\
    (e) & 172 & 219 & 281  \\
    (f) & 309 & 222 & 126 \\
    (g) & 729 & 932 & 530  \\
    (h) & 1750 & 560 & 716  \\
    (i) & 857 & 616 & 504 \\
    (j) & ... & 2190 & 2800  \\
	\hline
	\end{tabulary}\end{center}
	\caption{Analysis of the discovery potential for supersymmetry for the parameter space of Table~\ref{tab1}, using the 2 lepton same flavor (SF) selection criteria of Table~\ref{tab10}, where the minimum integrated luminosity needed for $5\sigma$ discovery is given in$\ifb$. Results from the different flavor (DF) SR {are not displayed} 
	 because of {their}  poor performance.}
	\label{tab11}
\end{table}

\begin{table}[H]
\begin{center}
\begin{tabulary}{0.85\textwidth}{lCCCCC}
\hline\hline\rule{0pt}{3ex}
Model & Leading SR & $\mathcal{L}$ (fb$^{-1}$) & Sub-leading SR & $\mathcal{L}$ (fb$^{-1}$)\\
\hline\rule{0pt}{3ex}
\!\!(g) & $0\ell4j$-A & 63 & $0\ell4j$-B & 147 \\
(e) & $0\ell4j$-A & 74 & $0\ell2j$-A & 83 \\
(a)& $1\ell2j$-C & 80 & $0\ell2j$-A & 86 \\
(b)& $2\ell1j$-SF-C & 93 & $0\ell2j$-A & 150\\
(c)& $0\ell2j$-A & 114 & $0\ell4j$-A & 749\\
(f)& $2\ell1j$-SF-C & 126 & $0\ell2j$-A & 173\\
(d)& $0\ell2j$-A & 174 & $2\ell1j$-SF-C & 243\\
(h)& $0\ell4j$-A & 337 & $0\ell2j$-A & 558\\
(i)& $2\ell1j$-SF-C & 504 & $2\ell1j$-SF-B & 616\\
(j)& $0\ell2j$-A & 771 & $2\ell1j$-SF-B & 2190 \\
\hline
\end{tabulary}\end{center}
\caption{
The overall minimum integrated luminosities needed for $5\sigma$ discovery using the
leading and the sub-leading signal regions for benchmarks of Table~\ref{tab1}, including all signal regions discussed. {The third column shows that all the benchmark  can be discovered with an integrated
luminosity below $1000$ fb$^{-1}$ which is significantly below the optimum integrated luminosity 
achievable at  the LHC}. }
\label{tab12}
\end{table}

\begin{table}[H]
\begin{center}
\begin{tabulary}{1.15\textwidth}{lCCCCCC}
\hline\hline\rule{0pt}{3ex}
Model & $\text{Br}(\tilde{G}\rightarrow\tilde{g}g)$ & $\text{Br}(\tilde{G}\rightarrow\tilde{\chi}_1^{\pm}W^{\mp})$ & $\text{Br}(\tilde{G}\rightarrow\tilde{\chi}_1^0\gamma)$ & $\text{Br}(\tilde{G}\rightarrow\tilde{\chi}_1^0 Z)$ & $\Gamma_{\tilde{G}}^{\text{two-body}}\times 10^{-24}$ & Lifetime  \\
&  &  &  &  & (GeV) & (s) \\
\hline\rule{0pt}{3ex}
\!\!(a)& 0.598 & 0.150 & 0.040 & 0.035 & 7.9 & 0.083 \\
(b)& 0.619 & 0.156 & 0.060 & 0.018 & 10.1 & 0.065 \\
(c)& 0.619 & 0.155 & 0.058 & 0.020 & 17.0 & 0.039 \\
(d)& 0.620 & 0.156 & 0.057 & 0.021 & 12.3 & 0.053 \\
(e)& 0.616 & 0.155 & 0.044 & 0.033 & 5.4 & 0.121\\
(f)& 0.616 & 0.155 & 0.043 & 0.034 & 6.6 & 0.099 \\
(g)& 0.614 & 0.155 & 0.041 & 0.036 & 3.6 & 0.183 \\ 
(h)& 0.618 & 0.155 & 0.038 & 0.039 & 14.1 & 0.047 \\
(i)& 0.617 & 0.155 & 0.041 & 0.037 & 4.4 & 0.151 \\
(j)& 0.617 & 0.155 & 0.036 & 0.042 & 8.9 & 0.074 \\ 
\hline
\end{tabulary}
\caption{Branching ratios of the leading decay channels of the gravitino, 
 the total two-body decay width and the lifetime of the gravitino for the benchmarks of Table~\ref{tab1}.}
\label{tab13}
\end{center}
\end{table}

\begin{table}[H]
\begin{center}
\begin{tabulary}{0.85\textwidth}{lCC}
\hline\hline\rule{0pt}{3ex}
Model & $\sigma^{\text{SI}}_{p,\chi^0_1}\times 10^{49}$ & $\sigma^{\text{SD}}_{p,\chi^0_1}\times 10^{47}$ \\
\hline\rule{0pt}{3ex}
\!\!(a)& 25.4 & 3.01 \\
(b)& 12.9 & 63.8 \\
(c)& 37.7 & 68.2 \\
(d)& 5.52 & 133 \\
(e)& 15.1 & 5.83 \\
(f)& 16.4 & 6.10 \\
(g)& 13.5 & 3.06 \\ 
(h)& 13.9 & 10.2 \\
(i)& 8.26 & 3.71 \\
(j)& 9.20 & 4.13 \\ 
\hline
\end{tabulary}
\caption{Proton--neutralino spin-independent ($\sigma^{\text{SI}}_{p,\chi^0_1}$) and spin-dependent ($\sigma^{\text{SD}}_{p,\chi^0_1}$) cross-sections in units of cm$^{-2}$ for the benchmarks  of Table~\ref{tab1}.}
\label{tab14}
\end{center}
\end{table}

\section{Figures}

\begin{figure}[H]
    \centering
	\includegraphics[width=0.45\textwidth]{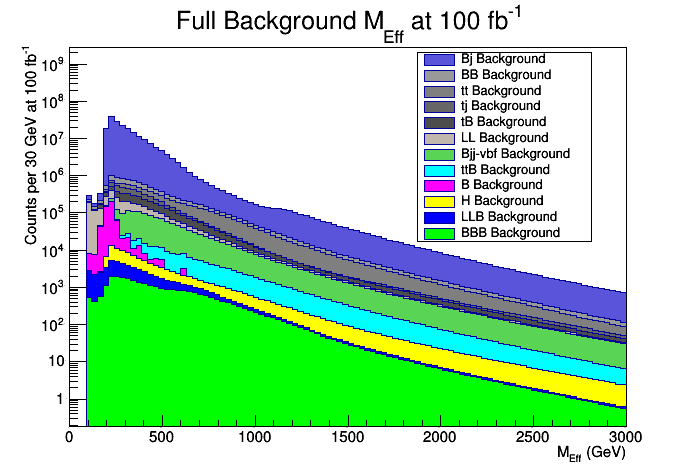}
    \includegraphics[width=0.45\textwidth]{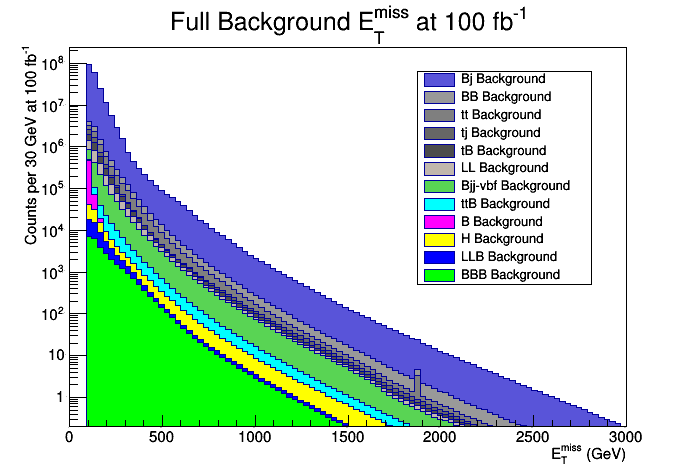}
    \caption{Full SNOWMASS standard model background~\cite{Avetisyan:2013onh} after triggering cuts and a cut of $E_T^{\text{miss}}\geq100\GeV$, broken into final states and scaled to $100\ifb$. The left panel gives $M_{\text{eff}}$(incl.) and the right panel gives $E_T^{\text{miss}}$. Individual data sets are labeled according to Eq. 1.}
	\label{fig:bgs}
\end{figure}

\begin{figure}[H]
    \centering
	\includegraphics[width=0.45\textwidth]{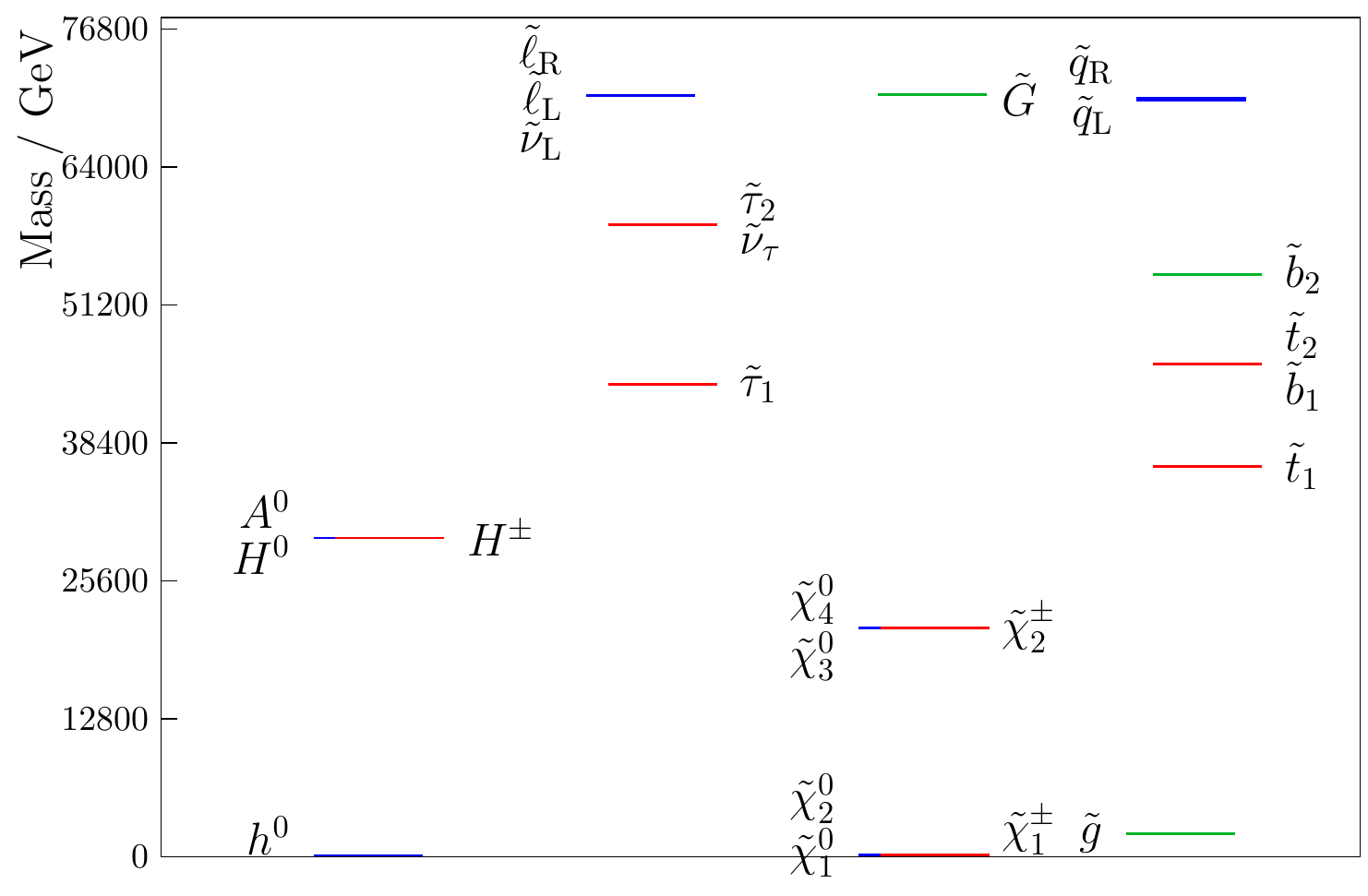}
    \includegraphics[width=0.45\textwidth]{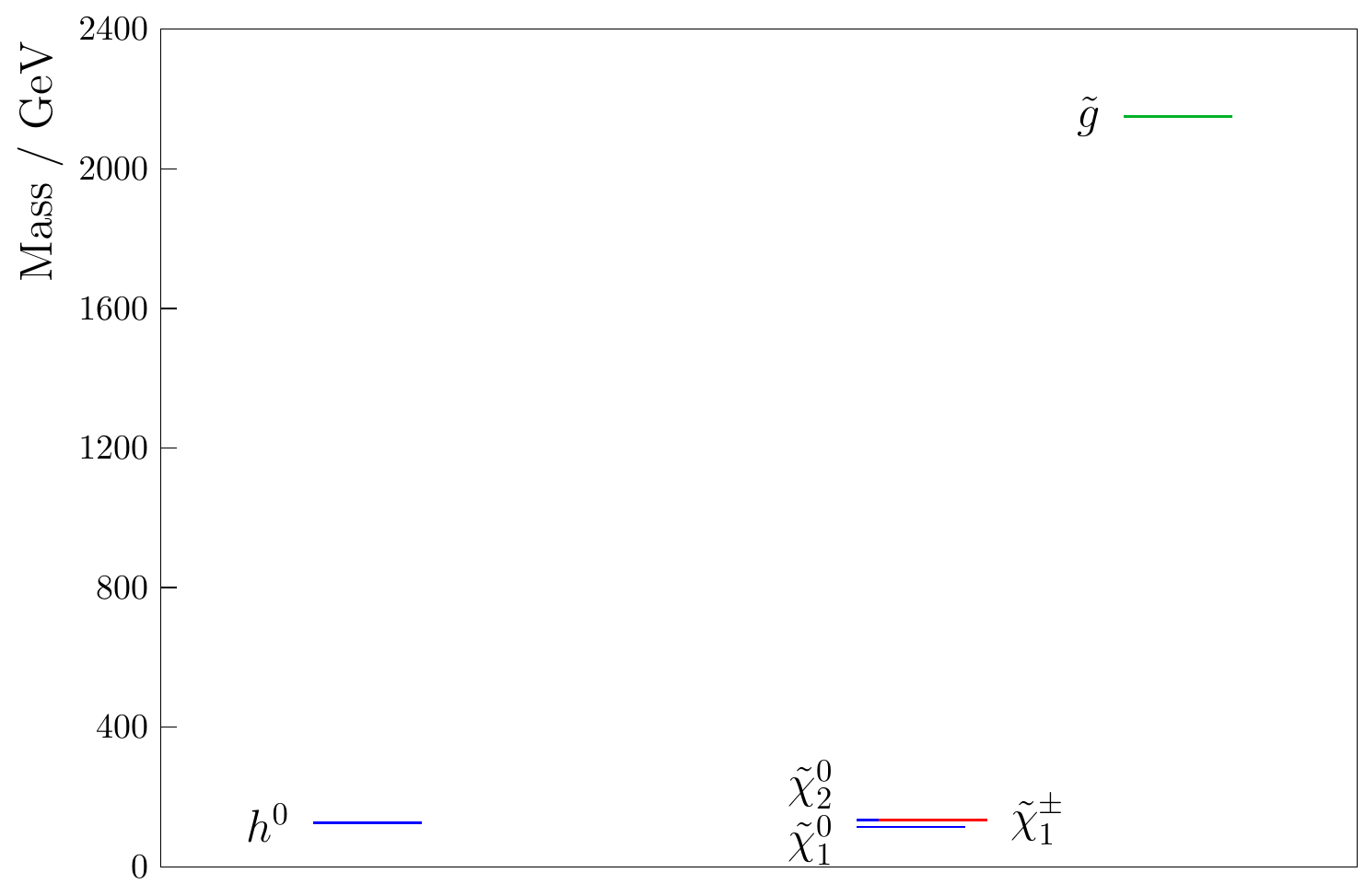}
    \caption{The sparticle spectrum for the benchmark  (a) of Table~\ref{tab1}. The figure in the left panel shows the entire spectrum with the heavy sfermions having mass at the order of $m_0$. The right panel shows the light spectrum which consists of the Higgs boson and the gauginos $\tilde \chi_1^0, \tilde\chi_2^0, \tilde\chi_1^{\pm}$ and  $\tilde g$.} 
	\label{fig1}
\end{figure}

\begin{figure}[htp]
\centering
\includegraphics[width=.4\textwidth]{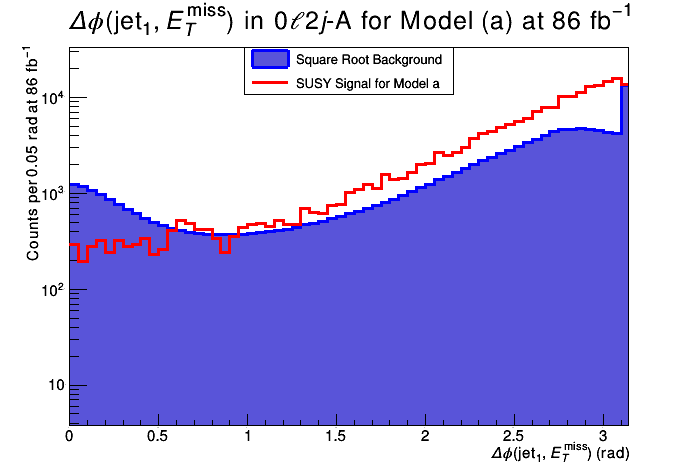}
\includegraphics[width=.4\textwidth]{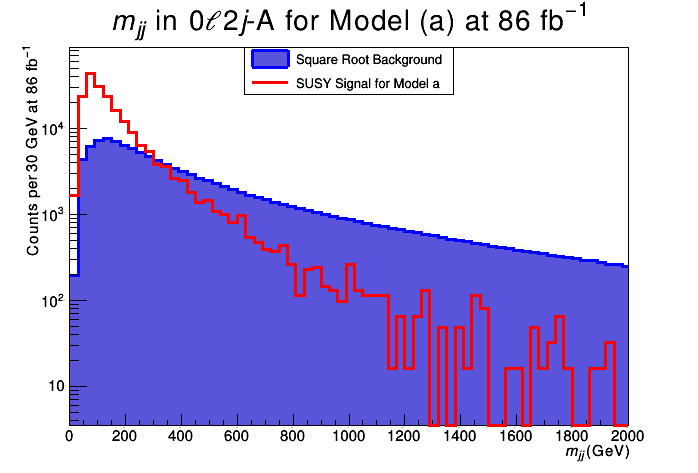}
\caption{Left panel: Distribution in $\Delta\phi(\text{jet}_1,E^{\text{miss}}_T)$ for the $0\ell2j$-A signal region defined in Table~\ref{tab6} for  the benchmark (a) of Table ~\ref{tab1}.
 Plotted is the number of counts for the SUSY signal per 0.05 rad and the square root of the total standard model SNOWMASS background. The analysis is done at 86 fb$^{-1}$ of integrated luminosity, which gives a 5$\sigma$ discovery in this signal region. Right panel: Distribution in the dijet invariant mass $m_{jj}$ where number of counts per 30 GeV is plotted for the same point as in the left panel.}
\label{fig2}
\end{figure}

\begin{figure}[htp]
\centering
\includegraphics[width=.4\textwidth]{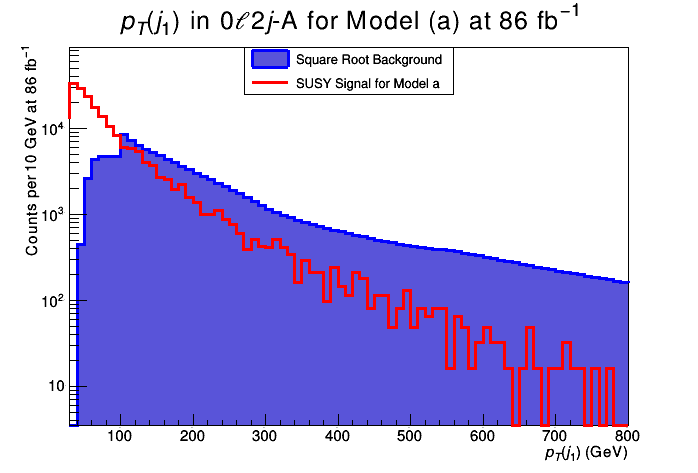}
\includegraphics[width=.4\textwidth]{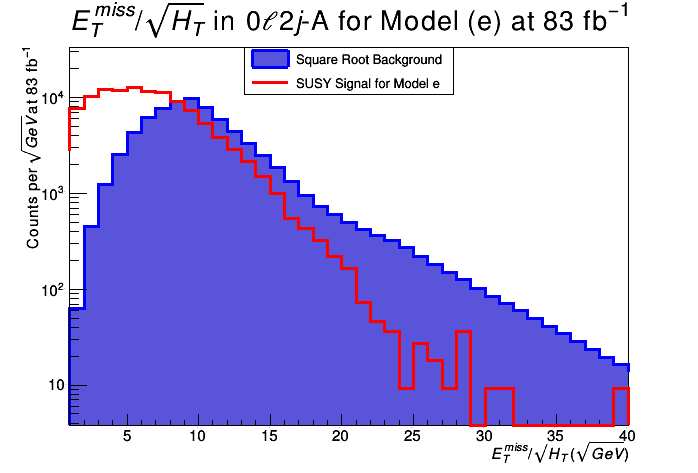}
\caption{Left panel: Distribution in $p_T(j_1)$ for the $0\ell2j$-A signal region defined in Table~\ref{tab6} for 
 benchmark (a) of Table ~\ref{tab1}. 
 Plotted is the number of counts for the SUSY signal per 10 GeV and the square root of the total standard model SNOWMASS background. The analysis is done at 86 fb$^{-1}$ of integrated luminosity, which gives a 5$\sigma$ discovery in this signal region. Right panel: Distribution in the variable $E^{\text{miss}}_T/\sqrt{H_T}$ where number of counts per $\sqrt{GeV}$ is plotted for
  benchmark (e) of Table ~\ref{tab1}at $83\ifb$.}
\label{fig3}
\end{figure}

\begin{figure}[htp]
\centering
\includegraphics[width=.4\textwidth]{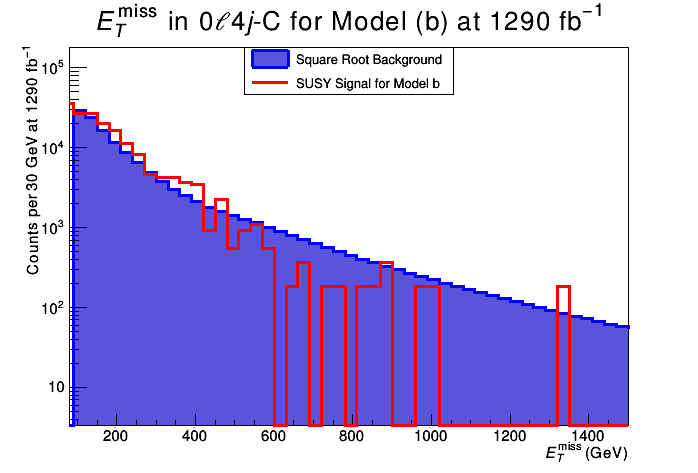}
\includegraphics[width=.4\textwidth]{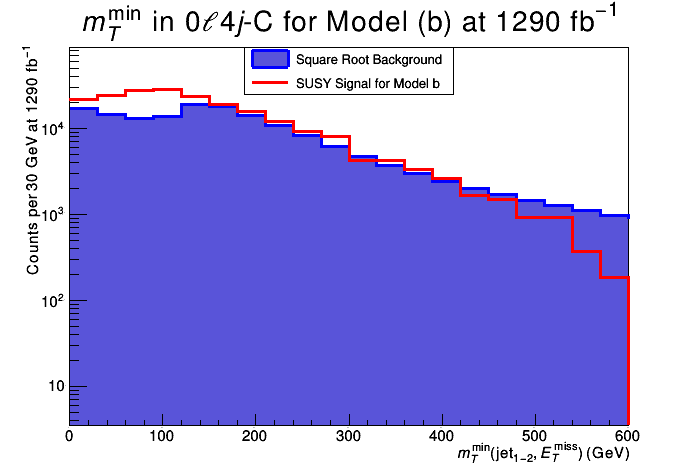}
\includegraphics[width=.4\textwidth]{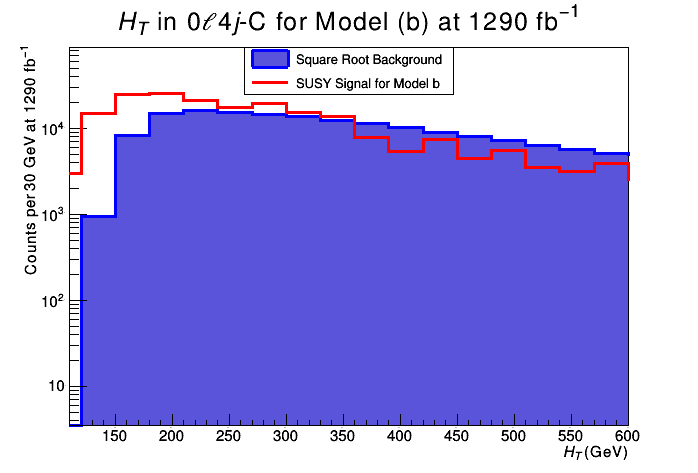}
\caption{Top left panel: Distribution in $E^{\text{miss}}_T$ for the $0\ell4j$-C signal region defined in Table~\ref{tab6} for 
benchmark (b) of Table ~\ref{tab1}.
 Plotted is the number of counts for the SUSY signal per 30 GeV and the square root of the total standard model SNOWMASS background. The analysis is done at 1290 fb$^{-1}$ of integrated luminosity, which gives a 5$\sigma$ discovery in this signal region. Top right panel: Same analysis as in adjacent panel but for the distribution in $m^{\text{min}}_T(\text{jet}_{1-2},E^{\text{miss}}_T)$. Bottom panel: Same analysis as in the top panels but for the distribution in $H_T$.}
\label{fig4}
\end{figure}

\begin{figure}[htp]
\centering
\includegraphics[width=.4\textwidth]{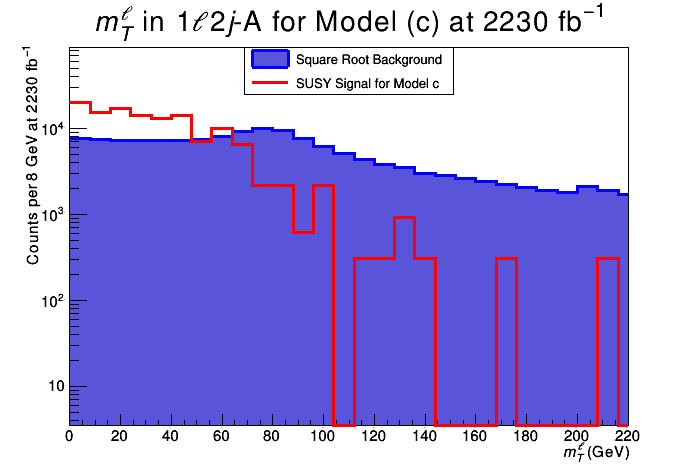}
\includegraphics[width=.4\textwidth]{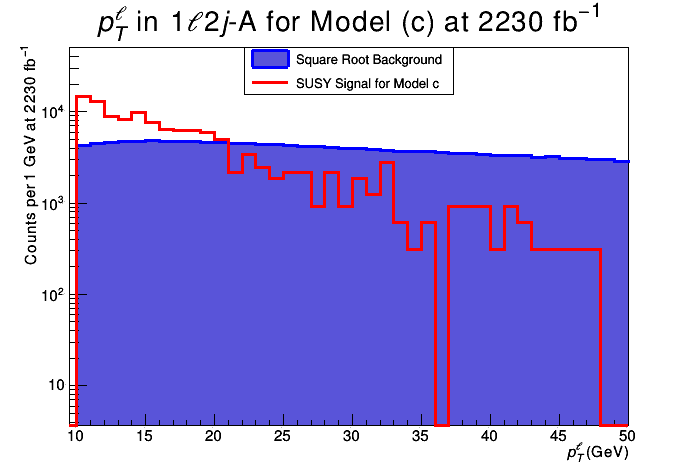}
\caption{Left panel: Distribution in $m^{\ell}_T$ for the $1\ell2j$-A signal region defined in Table~\ref{tab8} for
benchmark (c) of Table ~\ref{tab1}.
 Plotted is the number of counts for the SUSY signal per 8 GeV and the square root of the total standard model SNOWMASS background. The analysis is done at 2230 fb$^{-1}$ of integrated luminosity, which gives a 5$\sigma$ discovery in this signal region. Right panel: same analysis as in the left panel but for the distribution in $p^{\ell}_T$.}
\label{fig5}
\end{figure}

\begin{figure}[htp]
\centering
\includegraphics[width=.4\textwidth]{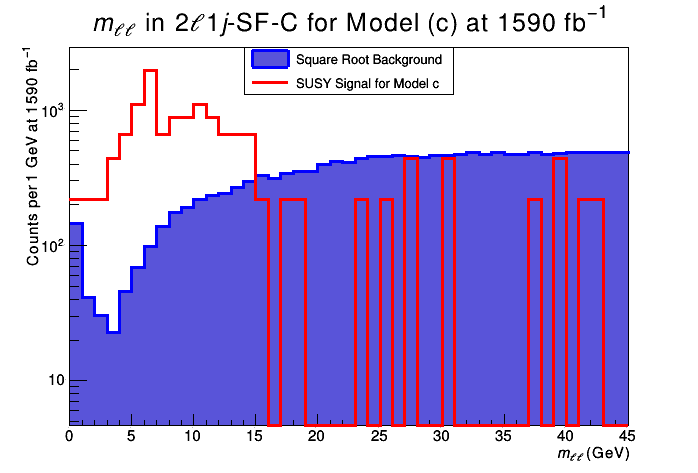}
\includegraphics[width=.4\textwidth]{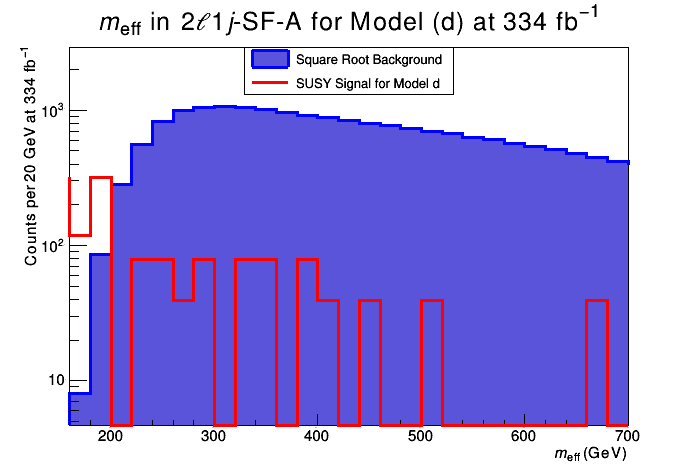}
\caption{Left panel: Distribution in the dilepton invariant mass, $m_{\ell\ell}$, for the $2\ell1j$-SF-C signal region defined in Table~\ref{tab10} for benchmark (c) of Table ~\ref{tab1}.
 Plotted is the number of counts for the SUSY signal per GeV and the square root of the total standard model SNOWMASS background. The analysis is done at 1590 fb$^{-1}$ of integrated luminosity, which gives a 5$\sigma$ discovery in this signal region. Right panel: Distribution in $m_{\rm eff}$ for the $2\ell1j$-SF-A signal region defined in Table~\ref{tab10} for 
 benchmark (d) of Table ~\ref{tab1}. 
 Plotted is the number of counts for the SUSY signal per 20 GeV and the square root of the total standard model SNOWMASS background. The analysis is done at 334 fb$^{-1}$ of integrated luminosity}
\label{fig6}
\end{figure}

\begin{figure}[H]
\begin{center}
	\includegraphics[width=0.8\textwidth]{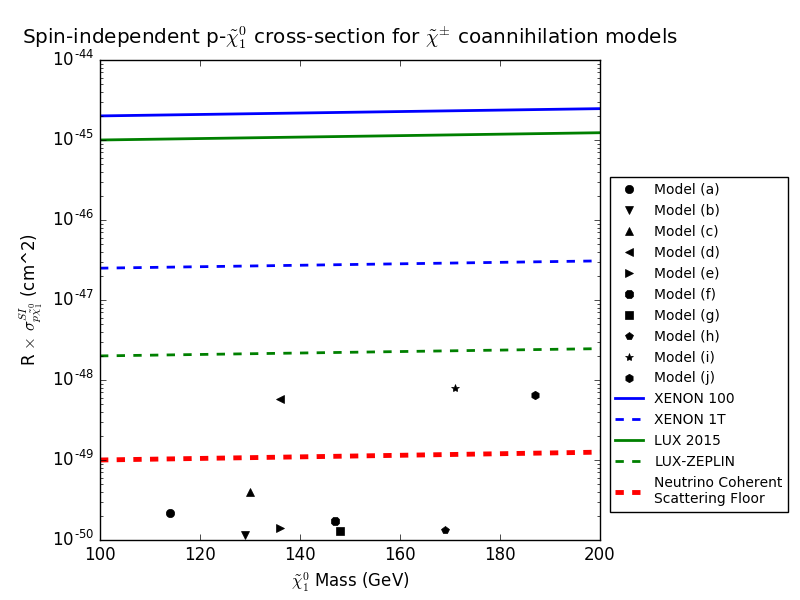}
\end{center}
    \caption{$R\times\sigma^{\text{SI}}_{p,\tilde{\chi}^0_1}$ $(R=\rho_{\tilde{\chi}^0_1}/\rho_c)$ for benchmarks     
     of Table~\ref{tab1} as a function of LSP mass displayed alongside the current and projected range of the XENON 
    and LUX experiments and the neutrino floor~\cite{Cushman:2013zza}. }
    \label{sigma}
\end{figure}

\end{document}